\documentclass[12pt, draftclsnofoot, onecolumn]{IEEEtran}
\usepackage{filecontents}
\usepackage{amsthm}
\usepackage[table,xcdraw]{xcolor}
\newtheorem{theorem}{\textbf{\text{Theorem}}}
\newtheorem{corollary}{Corollary}
\newtheorem{conjecture}{Conjecture}
\newtheorem{lemma}{Lemma}

\usepackage{epsfig}
\usepackage{times}
\usepackage{verbatim}
\usepackage{enumerate}
\usepackage{multicol}
\usepackage{afterpage}
\usepackage{wrapfig}
\usepackage{cite}
\usepackage{xfrac}
\usepackage{graphicx}
\usepackage[caption=false]{algorithm}
\usepackage{algorithmic}
\usepackage{ifmtarg}
\usepackage{amssymb}
\usepackage{amsmath}
\usepackage{color}
\usepackage{bbm}
\usepackage{epstopdf}
\usepackage{flushend}
\usepackage[font=small]{caption}
\usepackage{subcaption}
\newtheorem{definition}{\textbf{\text{Definition}}}
\newtheorem{assumption}{Assumption}
 \usepackage{multirow}
 \usepackage{stackengine}

\makeatletter
\newcommand*{\rom}[1]{\expandafter\@slowromancap\romannumeral #1@}
 \newcommand*{\bh}{\boldsymbol{h}}
  \newcommand*{\bH}{\boldsymbol{H}}
  \newcommand*{\bp}{\boldsymbol{p}}
    \newcommand*{\bu}{\boldsymbol{u}}
        \newcommand*{\bv}{\boldsymbol{v}}
        
        \usepackage[shortlabels]{enumitem}
\makeatother
\begin{document}
\title{Area Spectral Efficiency and SINR Scaling Laws in Multi-Antenna Cellular Networks}
	\author{
		\IEEEauthorblockN{\large  Ahmad AlAmmouri, Jeffrey G. Andrews, and Fran\c cois Baccelli}\\
		\thanks{The authors are with the Wireless Networking and Communications Group (WNCG), The University of Texas at Austin, Austin, TX 78712 USA. Email: \{alammouri@utexas.edu, jandrews@ece.utexas.edu, francois.baccelli@austin.utexas.edu\}. This work was supported 
in part by the National Science Foundation under Grant NSF-CCF-1514275
and in part by the Simons Foundation under Grant 197982. Part of this work has been submitted to ISIT 2020 \cite{Scaling_AlAmmouri20}. Last revised \today.}}
	
	\maketitle
	\begin{abstract}
	We study the scaling laws of the  signal-to-interference-plus-noise ratio (SINR) and area spectral efficiency (ASE) in multi-antenna cellular networks, where the number of antennas scales with the base station (BS) spatial density $\lambda$. We start with the MISO case having $N_t(\lambda)$ transmit antennas and a single receive antenna and prove that the average SINR scales as $\sfrac{N_t(\lambda)}{\lambda}$ and the average ASE scales as $\lambda\log\left(1+\sfrac{N_t(\lambda)}{\lambda}\right)$. For the MIMO case with single-stream eigenbeamforming and $N_r(\lambda) \leq N_t(\lambda)$ receive antennas, we prove that the scaling laws of the conditional SINR and ASE are  exactly the same as the MISO case, i.e. not dependent on $N_r(\lambda)$. We also show that coordinated beamforming amongst $K\leq N_t(\lambda)$ neighboring BSs does not improve the scaling laws regardless of $K$. From a system design perspective, our results suggest that deploying multi-antenna BSs can help maintain the per-user throughput and the linear increase in the ASE with BS density, while the number of antennas at the user equipment and the use of BS cooperation do not matter much. 
	\end{abstract}
\begin{IEEEkeywords}
Area spectral efficiency, cellular networks, MIMO, stochastic geometry, ultradense networks.
\end{IEEEkeywords}
\section{Introduction}
Earlier academic works \cite{A_Andrews11} and several decades of industry deployments indicate that area spectral efficiency (ASE) -- the network sum throughput per unit area -- increases about linearly with network densification, namely the base station (BS) density,  In contrast, we recently showed that under natural assumptions on the network and signal propagation models, the signal-to-interference-plus-noise ratio (SINR) degrades to zero and the ASE saturates to a finite constant in the limit of very dense networks \cite{A_AlAmmouri19}. Practically, this result means that densifying the network beyond a certain point actually sacrifices the per-user performance, and unfortunately the per-user rate drops asymptotically to zero, a result corroborated by other recent works on ultra-dense networks \cite{Downlink_Zhang15,Performance_Ding17,Performance_Nguyen17,SINR_AlAmmouri17}.  All those works were on single antenna transmission and reception.  In this work, we study whether deploying multi-antenna BSs and/or users' equipment (UEs) can improve the scaling laws of the SINR and the ASE in cellular networks. The motivation is that multi-antenna systems typically improve the per-user SINR by increasing the channel (array) gain and/or mitigating the network interference. 

 \subsection{Related Work}
The history of studying the scaling laws in decentralized (ad hoc) wireless networks is largely credited to \cite{The_Gupta00}, which showed that despite the intractability of the  exact network capacity, one can still characterize how the per-node throughput scales with the number of nodes.  A subsequent approach that is able to more accurately quantify the SINR and spatial throughput of decentralized wireless networks relies on tools from stochastic geometry \cite{Stochastic_Baccelli10_2,Stochastic_Haenggi12}, as well-summarized in \cite{Stochastic_ElSawy13,Transmission_Weber11}. Particularly relevant to the multi-antenna case are  \cite{Transmission_Hunter08,Open_Louie11,Transmission_Vaze12,Multi_Jindal10,Spatial_Huang12,Spectral_Lee16,Scaling_Lee18}. In \cite{Transmission_Hunter08,Open_Louie11,Spatial_Huang12,Multi_Jindal10,Transmission_Vaze12}, the authors studied the transmission capacity in ad hoc networks, which is the network sum throughput per unit area, assuming no link adaptation and with outage constraints. Some of the key results in these works are: assuming a single data stream, the authors in \cite{Transmission_Hunter08} showed that dynamic beamforming yields a better scaling law than selection combining and space-time orthogonal coding, while in \cite{Open_Louie11}, the authors showed that  spatial multiplexing, i.e., multiple data streams, can improve the transmission capacity in certain scenarios, and in \cite{Transmission_Vaze12}, it was shown that a single data stream is optimal in terms of the transmission capacity when the receiver employs interference cancellation. In \cite{Multi_Jindal10}, it was shown that even with single-antenna transmitters, interference cancellation provides a linear scaling of the transmission capacity with the nodes density and these results were extended to the MIMO settings in \cite{Spatial_Huang12}.
 
 In \cite{Spectral_Lee16,Scaling_Lee18}, the authors studied the scaling laws of multi-antenna ad hoc networks in terms  of the ASE, which is the network sum throughput per unit area with full link adaption and no outage constraint. Hence, the ASE upper bounds the transmission capacity and is more relevant to modern wireless networks, where link adaptation is implemented \cite{Are_Lozano12}.  The three key results in \cite{Spectral_Lee16}, which focuses on SIMO networks, are: the ASE asymptotically drops to zero in the single-antenna case, super-linear scaling of the number of antennas is required to maintain a linear scaling of the ASE, and nodes cooperation improves the ASE scaling law. This model was extended to the MIMO settings in \cite{Scaling_Lee18} and the authors showed that spatial multiplexing can improve the ASE scaling laws in certain settings. Hence, there is potential to improve the scaling laws of the ASE by increasing the number of antennas. Interestingly, the scaling laws in cellular networks are different from the ones in \cite{Spectral_Lee16}, as we show in this work.
 
 In a cellular network, the authors in \cite{A_Andrews11} studied the performance of single-antenna BSs and users, and the key result derived was that the ASE scales linearly with the BS density, assuming the standard power-law path loss model, where the single power drops by $r^{-\eta}$ over a distance $r$ with $\eta>2$. This model has been extended to many other network configurations as well-summarized in \cite{Modeling_ElSawy16,A_Andrews16,Stochastic_Bartek18}.  In the multi-antenna context, the works \cite{Downlink_Dhillon13,A_Renzo14,Stochastic_Renzo15,A_Afify16,Analysis_Tanbourgi15} studied different performance metrics of MIMO cellular networks, also under the assumption of the standard power-law path loss model. For example, ordering results for the different MIMO techniques were derived in \cite{Downlink_Dhillon13} in terms of the coverage probability, expressions for the bit error probability were derived in \cite{A_Renzo14,Stochastic_Renzo15,A_Afify16}, and in \cite{Analysis_Tanbourgi15}, the case of   orthogonal space–time block
coding at the BSs and maximal-ratio combining at the users was studied. 

Using the power-law model has been a common practice in studying the performance of wireless networks since it leads to more tractable analysis and has a very long history \cite{Empirical_Hata80}. Nevertheless, it is known that this model is inaccurate for short distances, due to the singularity at the origin, and, at least in the single-antenna case, the scaling laws derived under this model can be misleading as shown in \cite{A_AlAmmouri19,Downlink_Zhang15,Performance_Ding17,Performance_Nguyen17,SINR_AlAmmouri17}. Precisely, the linear scaling law derived in \cite{A_Andrews11} vanishes once  a more physically feasible path loss model is considered.
 
\subsection{Summary of Contributions}

We assume that the BSs are spatially distributed as a homogeneous Poisson point process (HPPP) with density $\lambda$, the use of any physically feasible path loss model, and independent and identically distributed (i.i.d.) circularly symmetric complex Gaussian channels between the different transmit and receive antennas. For the MISO case with $N_t(\lambda)$ transmit antennas, we prove that the average SINR scales as $\sfrac{N_t(\lambda)}{\lambda}$ and the average ASE scales as $\lambda\log\left(1+\sfrac{N_t(\lambda)}{\lambda}\right)$. Then we generalize the results for the MIMO case with eigenbeamforming, a single stream of data, and $N_r(\lambda)\leq N_t(\lambda)$ receive antennas. We prove that the scaling laws of the conditional SINR, i.e., conditional on the network geometry and channel gains, and the conditional ASE are agnostic to  $N_r(\lambda)$ and scale exactly the same as the MISO case. Then we show that BSs cooperation through coordinated beamforming does not improve the scaling laws of the ASE.  Hence, under the above i.i.d. assumption, one can maintain a non-zero per-user throughput in dense networks if the number of antennas is properly scaled with the BS density. Consequently, a desired growth of the ASE can be guaranteed asymptotically.

Note that the two major differences between our model and the one in \cite{Scaling_Lee18,Spectral_Lee16} are: ($i$) we consider a cellular network, while an ad hoc network was studied in \cite{Scaling_Lee18,Spectral_Lee16} and ($ii$) the unbounded power-law path loss was assumed in \cite{Scaling_Lee18,Spectral_Lee16}, while we consider the general class of physically feasible path loss models. In the last section of this work, we modify the network model we consider to match the ad hoc model considered in \cite{Scaling_Lee18,Spectral_Lee16} and we show that scaling laws for the ad hoc case match the ones we derived for the cellular case. Hence, the differences between our scaling laws and the ones in \cite{Scaling_Lee18,Spectral_Lee16} are only due to the path loss model, and in general, the scaling laws  we derive for the ad hoc case are more optimistic compared to \cite{Scaling_Lee18,Spectral_Lee16}. In terms of the methodology of the analysis, this work follows the same approach we followed in \cite{A_AlAmmouri19}, where we focus on a wide class of path loss functions, then use tools from probability theory and stochastic geometry, such as the law of large numbers and the infinite divisibility of PPPs, to derive the scaling laws for the desired performance metrics. This approach is different from the one used   in \cite{Scaling_Lee18,Spectral_Lee16}, which relies on computing the exact expressions of the desired performance metrics, typically involve multiple integrals,  then rely on bounds to derive the scaling laws.

 \subsection{Paper Organization}
  In Section \ref{Sec:SysMod}, we present the system model and the basic assumptions. In Section \ref{Sec:Meth}, we present the methodology of the analysis along with the performance metrics. Section \ref{Sec:MISO} focuses on multi-antenna BSs and single-antenna users, i.e., MISO network, and the SIMO case, with single-antenna BSs and multi-antenna users is discussed in Section \ref{Sec:SIMO}. The case of multi-antenna BSs and users is studied in Section \ref{Sec:MIMO}. In Section \ref{Sec:MISOCoor}, we focus on MISO networks with BS cooperation through coordinated beamforming. Discussions, conclusion, and future works are presented in Section \ref{Sec:Discussion}.

	\begin{table}[t]
	\centering
	\caption{Notation.}
	\label{tb:Notation}
	{%
		\begin{tabular}{|l|l|}
			\hline
			\rowcolor[HTML]{EFEFEE} 
			Notation & Definition \\ \hline
			$\lambda$ & The spatial density of the BSs. \\ \hline
			$\Phi$ & A Poisson point process with density $\lambda$.\\\hline
			$L(r)$ & The average channel gain over a distance $r$. \\ \hline
			$N_t(\lambda)$ & The number of transmit antennas.\\ \hline
			$N_r(\lambda)$ & The number of receive antennas.\\ \hline
			$B(x,R)$ &  A disk centered at $x$ with radius $R$. \\ \hline
			$L_0$ &  The transmit power of the BSs, i.e., $L_0=L(0)$. \\ \hline
			$\gamma$ &  $\gamma:=\int\limits_{0}^{\infty} r L(r) dr.$ \\ \hline
			$\bH_{i,j}$ &  The channel between the $i^{\rm th}$ BS and the $j^{\rm th}$ user. \\ \hline
			$\bp_{i}$ &  The precoding vector used by the $i^{\rm th}$ BS.\\ \hline
				$\bu_{i}$ &  The combining vector used by the $i^{\rm th}$ user.\\ \hline
			$s_i$ &  The transmitted symbol from the $i^{\rm th}$ BS.\\ \hline
			$\sigma^2$ & The average noise power in Watts. \\ \hline
	\end{tabular}}
\end{table}

\section{System Model}\label{Sec:SysMod}
In this section, we discuss the main assumptions we have on the network model, the propagation model, the beamforming architecture, and the antenna configurations. The notation is summarized in Table \ref{tb:Notation}.

{\bf Network Model:} We consider a single-tier downlink cellular network where the BSs are spatially distributed as a two-dimensional HPPP $\Phi$ with intensity $\lambda$ \cite{Stochastic_Baccelli10_2}. Users are spatially distributed according to an independent stationary point process, with intensity $\lambda_u \gg \lambda$, such that each BS has at least one user to serve with probability one. Each BS schedules its users on orthogonal resource blocks such that one user is associated with every BS in a given resource block. Hence, users do not suffer from intra-cell interference, but they are still affected by interference from other cells. Based on this, the intensity of active users in a given resource block is equal to $\lambda$. Users are assumed to connect to their closest BS, i.e., the BS with the highest average received power and all BSs are assumed to transmit with a unit power distributed across their antennas. Hence, our network model matches the one studied in \cite{A_Andrews11}.

{\bf Propagation Model:} The large-scale channel gain is assumed to be captured by the function $L:\mathbb{R}_{+}\rightarrow \mathbb{R}_{+}$, i.e., $L^{-1}(\cdot)$ is the path loss. We focus on the class of physically feasible path loss models introduced in \cite{A_AlAmmouri19}, which is defined as follows.
\begin{definition}\label{Def:PLM}(Physically feasible path loss)
	Physically feasible path loss models are the family of path loss functions $L(r)$ with the following properties 
	\begin{enumerate}
		\item $L_0=L(0) \in \mathbb{R}_{+}$.
		\item $L(r)\leq L_0 \ \forall r\in \mathbb{R}_{+}$.
		\item $\gamma=\int\limits_{0}^{\infty} r L(r)dr \in \mathbb{R^{*}_{+}}$.
	\end{enumerate}
	\end{definition}
The first requirement translates to having a finite BS transmit power; the second ensures that the average received power is less than the transmit power, and the third guarantees that the sum of received powers from all BSs is almost surely (a.s.) finite  at any location in the network. Note that although the third property follows from Campbell's theorem \cite{Stochastic_Baccelli10_2}, which means that this condition by itself translates to having a finite interference in the mean sense, but it is also necessary and sufficient for the finiteness of the interference in the almost sure sense according to \cite[Theorem 4.6]{Stochastic_Haenggi12} due to the boundedness of $L(\cdot)$. For more information about this class of path loss models, refer to \cite{A_AlAmmouri19}.  

The bounded single-slope, the bounded multi-slope \cite{Downlink_Zhang15}, and the stretched exponential \cite{SINR_AlAmmouri17} path loss models in addition to the path loss models used in 3GPP standards \cite{3GPP2017} for the entire range of    0.5    to    100   GHz bands, are all included in this class of models. However, due to the singularity at $r=0$, the power-law model, i.e., $r^{-\eta}$, is not included in this class, since it violates the second property of the physically feasible path loss models.

To obtain some of the results in this work, we need additional assumptions on the path loss model to maintain analytical tractability. These assumptions are summarized in the following.

\begin{assumption}
	The path loss function satisfies the following: $\exists \  r_0 \in {R}_{+}$,  $\zeta \in \mathbb{R}_{+}^{*}$, and a differentiable decreasing function $\tilde{L}: [r_0,\infty)\rightarrow \mathbb{R}_{+}$ such that:
	\begin{enumerate}
		\item $\tilde{L}(r)\leq L(r), \ \forall r \in [r_0,\infty)$.
		\item $\frac{r\tilde{L}(r)}{-\tilde{L}^{'}(r)} \geq \zeta ,\  \forall r \in [r_0,\infty)$.
		\item $\int\limits_{r_0}^{\infty} \frac{r}{\tilde{L}(r)^2} e^{- \pi \lambda_0 r^2}dr$ is finite for all $\lambda_0>\lambda_c\in \mathbb{R_{+}}$.
	\end{enumerate}
\end{assumption}

The conditions in Assumption 1 do not have a clear intuitive meaning, but the path loss models we mentioned previously - the bounded single-slope, the bounded multi-slope, the stretched exponential, and the 3GPP path loss models - all satisfy the three conditions in Assumption 1 \cite{A_AlAmmouri19}. An example of a path loss function that is physically feasible but does not satisfy the previous conditions is a bounded function with a bounded support, i.e., $L(r)=0, \ \forall r>r_0$ and $L(r)\leq L_0, \ \forall r\leq r_0$ for some $r_0,L_0\in \mathbb{R}_{+}$.

All small-scale fading variables between any two nodes are assumed to be i.i.d. and independent of the locations of the nodes. We focus on the digital beamforming architecture, where each antenna is connected through a separate RF chain, and hence, the nodes have direct access to the channel seen by each antenna. The channel, i.e., the small-scale fading, between any transmit antenna and receive antenna, is assumed to follow i.i.d.  circularly symmetric complex Gaussian random variables with zero mean and unit variance,  which inherently means we assume a rich scattering environment with the proper antenna spacing \cite{Foundations_Heath18}. This assumption is questionable when the network utilizes frequency bands in the mmWave and THz bands since the channels are known to be spatially sparse with a few dominant paths \cite{Millimeter_Rappaport14}. Hence, our model is more suitable for traditional cellular bands and we postpone considering spatially sparse channels for a future work. Note that we assume an environment which is rich enough with scatters such that the i.i.d. assumption is valid regardless of the number of antennas, which could be huge. 

{\bf Antenna  Configurations:}
The BSs are equipped with $N_t(\lambda)$ antennas and users with $N_r(\lambda)$ antennas. More specifically, we consider the following four scenarios:
\begin{itemize}
    \item {\it Scenario A}: MISO networks, where we have multi-antenna BSs and single-antenna users with no BS cooperation.
    \item {\it Scenario B}: SIMO networks, where we have single-antenna BSs and multi-antenna users with no interference cancellation.
     \item {\it Scenario C}: MIMO networks, where we have multi-antenna BSs and multi-antenna users with no BS cooperation nor interference cancellation. In this case, only a single stream of data is allowed, regardless of the number of transmit and receive antennas, i.e., no data multiplexing. 
     \item {\it Scenario D}: MISO networks, where we have multi-antenna BSs and single-antenna users, but with BS cooperation through coordinated beamforming.
\end{itemize}

In all scenarios, we assume that each BS and its user have perfect knowledge of the channel between them. In the first three scenarios, each BS only knows the channel to its own user with no information shared between the BSs, while in the last scenario, BSs cooperate to enhance the performance of their users. Furthermore, in the MISO and the MIMO cases, $N_t(\lambda)$ is assumed to positively scale with $\lambda$, i.e., it is non-decreasing with $\lim\limits_{\lambda \rightarrow \infty}N_t(\lambda)=\infty$, while in the SIMO case, $N_r(\lambda)$ is assumed to positively scale with $\lambda$.  In the next section, we present the methodology of the analysis we follow to prove the scaling laws for the four scenarios of interest.

\section{Methodology of Analysis}\label{Sec:Meth}
We consider the performance of a user located at the origin. According to Slivnyak's theorem \cite{Stochastic_Baccelli10_2}, there is no loss of generality in this assumption, and the evaluated performance represents the average performance for all users in the network. The received signal at the tagged user assuming a serving distance of $r_0$ is
\begin{align}
    y_0&=\sqrt{L(r_0)}\bu^{*}_{0}\bH_{0,0}\bp_{0} s_0 +\sum\limits_{r_i\in \Phi \setminus B(0,r_0)} \sqrt{L(r_i)}\bu^{*}_{0}\bH_{i,0}\bp_{i} s_{i}+\bu^{*}_{0}n_0,
\end{align}
where $\bH_{i,j}\in \mathbb{C}^{N_r\times N_t}$ is the channel between the $i^{\rm th}$ BS and the $j^{\rm th}$ user, $\bp_{i}\in \mathbb{C}^{N_t\times1}$ is the precoding (beamforming) vector of the $i^{\rm th}$ BS, $\bu_{i}\in \mathbb{C}^{N_r\times1}$ is the combining vector used by the $i^{\rm th}$ user, $s_i$ is the transmitted symbol from the $i^{\rm th}$ BS, $n_0$ is the zero-mean additive white Gaussian noise with variance $\sigma^2$, and $B(0,r_0)$ is a ball centered at the origin with radius $r_0$. Hence, $\Phi \setminus B(0,r_0)$ is the set of interfering BSs. Note that for simplicity, we assume that $\Phi$ is the ordered set of distances between the origin and the BSs, hence, $r_0$ is the closest point to the origin. Also, note that the users are assumed to be ordered such that the $i^{\rm th}$ user is connected to the $i^{\rm th}$ BS, which is possible since each BS serves exactly one user on the considered resource block.

The transmitted symbols from the BSs are assumed to be i.i.d. with zero mean and unit energy. We further adopt the  Gaussian signaling approximation \cite{A_Afify16}, where the interfering symbols are assumed to be drawn from i.i.d. complex Gaussian random variables with zero mean and unit variance, i.e., $s_i\sim\mathcal{CN}(0,1), \ \forall i \in \{1, 2, \cdots\}$. The accuracy of the Gaussian codebook approximation for interfering symbols from several constellation types has been verified in \cite{A_Afify16,The_Afify15,Influence_Giorgetti05}. 
Hence, by conditioning on the network geometry, the channel gains, and the precoding/combining vectors, the interference is conditionally a Gaussian random variable.  Under such a conditioning and by averaging over the transmitted symbols, the desired signal power is $L(r_0) |\bu^{*}_{0}\bH_{0,0}\bp_{0}|^2$ and the interference-plus-noise power is $\sum\limits_{r_i\in \Phi \setminus B(0,r_0)}L(r_i)|\bu^{*}_{0}\bH_{i,0}\bp_{i}|^2+||\bu_0||_{2}^{2}\sigma^2$, since the transmitted symbols are assumed to be i.i.d and the interference is treated as noise. Hence, the conditional SINR is represented by
\begin{align}\label{Eq:SINR_gen}
 {\rm SINR}(\lambda)&=\frac{L(r_0) |\bu^{*}_{0}\bH_{0,0}\bp_{0}|^2}{\sum\limits_{r_i\in \Phi \setminus B(0,r_0)}L(r_i)|\bu^{*}_{0}\bH_{i,0}\bp_{i}|^2+||\bu_0||_{2}^{2}\sigma^2},
\end{align}
where the dependency on $\lambda$ is captured through the distribution of the serving distance $r_0$, the density of interfering BSs, and the number of antennas. Note that the serving distance has a Rayleigh  probability density function (PDF) $f_{R}(r_0)=2 \pi \lambda r_0 e^{-\pi \lambda r_0^2}$ \cite{A_Andrews11}.

Both the BS and the user design their precoding and combining vectors, respectively, to maximize the SNR at the user. Under the assumption that the elements of $\bH$ are drawn from i.i.d. complex Gaussian random variables with zero mean and unit variance, the BS (user) uses the right (left) singular vector corresponding to the maximum eigenvalue of the matrix $\bH$ as its beamforming (combining) vector, which is referred to as eigenbeamforming.  Based on \cite{Transmission_Hunter08}, the SINR in this case can be expressed as
\begin{align}\label{Eq:SINR_MIMO2}
   {\rm SINR}(\lambda)&=\frac{L(r_0) \phi_{0}^2}{\sum\limits_{r_i\in \Phi \setminus B(0,r_0)}L(r_i)g_i+\sigma^2},
\end{align}
where $\phi_{0}$ is the maximum singular value of the matrix $H_{0,0}$ and $g_i, \forall i \in \{1,2, \cdots\},$ are i.i.d. unit mean exponential random variables independent of $\phi_{0}$. 

The second performance measure we consider is the  ASE \cite{Area_Alouini99}, which represents the sum throughput for all users per unit area. Given our system model, we define the conditional ASE  as \cite{A_AlAmmouri19}
\begin{align}\label{Eq:ASEGen}
    \mathcal{E}(\lambda) =\lambda \log_2(1+ {\rm SINR}(\lambda)) ,
\end{align}
in bps/Hz/m$^{2}$. Note that the average SINR and the average ASE can be found by averaging \eqref{Eq:SINR_gen} and \eqref{Eq:ASEGen}, respectively, over all channel realizations,  precoding vectors, and network configurations. The average SINR captures the per-user performance, while the average ASE represents the network sum throughput per unit area.\footnote{Note that the physical meaning of the conditional SINR is clear as  we explained. However, the conditional ASE in the way we defined it doesn't have a clear physical meaning, since it is the product of the average number of links per-unit area, i.e., $\lambda$, and the conditional per-user throughput. Hence, the average ASE is the correct performance metric to study.}  In terms of scaling laws, Fatou's lemma \cite{Real_Royden88} is very helpful since it shows that $\lim\limits_{\lambda \rightarrow \infty}\mathbb{E}[{\rm SINR}(\lambda)]\geq \mathbb{E}[\lim\limits_{\lambda \rightarrow \infty}{\rm SINR}(\lambda)]$ and $\lim\limits_{\lambda \rightarrow \infty}\mathbb{E}[\mathcal{E}(\lambda)]\geq \mathbb{E}[\lim\limits_{\lambda \rightarrow \infty}\mathcal{E}(\lambda)]$. Hence, by deriving the scaling laws for the conditional SINR (ASE), we can derive lower bounds on the scaling laws of the average SINR (ASE). Motivated by this, in the following sections, we start by deriving the scaling laws of the conditional SINR (ASE) first, and then move to the average SINR (ASE), where we utilize the conditions summarized in Assumption 1 to prove that the average SINR (ASE) has the same scaling laws as conditional SINR (ASE). Another fundamental lemma that we rely on is given next.\footnote{A special case of this lemma was implicitly proved in \cite{A_AlAmmouri19}.}
\begin{lemma}\label{Lem:Thm1_1}
Let $L(\cdot)$ be a general physically feasible path loss model, $g_i, i \in \{1,2, \cdots \},$ be a sequence of i.i.d. random variables with unit mean, $\Phi$ be a HPPP with intensity $\lambda$, and $r_n$ be the distance from the origin to the $n^{\rm th}$ closest points in $\Phi$, then for all fixed $n$,
\begin{align}
    \lim\limits_{\lambda \rightarrow \infty} \frac{1}{\lambda} \sum\limits_{r_i \in \Phi \setminus B(0,r_n)}L(r_i) g_i= 2 \pi \gamma \ {\rm a.s.}
\end{align}
where $\gamma=\int\limits_{0}^{\infty} r L(r) {\rm d}r$.
\begin{proof}
Let $\lambda=k \lambda_0$, where $k \in \mathbb{Z}_{+}$ and $\lambda_0 \in \mathbb{R}^{*}_{+}$, and $\tilde{\Phi}$ be a PPP with intensity $k \lambda_0$. Then,
\begin{align}
  \lim\limits_{k \rightarrow \infty} \frac{1}{k\lambda_0} \sum\limits_{r_i \in \tilde{\Phi} \setminus B(0,r_n)}L(r_i) g_i&=   \lim\limits_{k \rightarrow \infty}\left( \frac{1}{k\lambda_0}\sum\limits_{r_i\in \tilde{\Phi}}g_i L(r_i)-\frac{1}{k\lambda_0}\sum\limits_{j=0}^{n-1}g_j L(r_j)\right)\label{eq:Lem1_1}\\
  &=\lim\limits_{k \rightarrow \infty} \frac{1}{k\lambda_0}\sum\limits_{r_i\in \tilde{\Phi}}g_i L(r_i)\label{eq:Lem1_2}\\
  &=\lim\limits_{k \rightarrow \infty}\frac{1}{k\lambda_0}\sum\limits_{j=1}^{k}\sum\limits_{r_{j,i}\in \Psi_{n}}g_{j,i} L(r_{j,i})\label{eq:Lem1_3}\\
  &=\frac{1}{\lambda_0}\mathbb{E}\left[\sum\limits_{r_{0,i}\in \Psi_{0}}g_{0,i} L(r_{0,i})\right]\label{eq:Lem1_4}\\
  &=2 \pi \int\limits_{0}^{\infty}r L(r)dr=2 \pi \gamma,\label{eq:Lem1_5}
\end{align}
where \eqref{eq:Lem1_1} follows by  adding and subtracting the interference from the $n^{\rm th}$ closest points to the origin, \eqref{eq:Lem1_2} holds since $\frac{\sum\limits_{j=0}^{n-1}g_j L(r_j)}{k\lambda_0}\leq \frac{L_0 \sum\limits_{j=0}^{n-1}g_j }{k\lambda_0}$ which approaches zero a.s. as $k \rightarrow \infty$, given that $n$ is finite, \eqref{eq:Lem1_3} follows using the superposition property of PPPs \cite{Stochastic_Baccelli10_2}, where $\Psi_j, \ \forall j \in \{1,2, \cdots, k\}$ are i.i.d. PPPs with density $\lambda_0$,  \eqref{eq:Lem1_4} follows by using the law of large numbers, and \eqref{eq:Lem1_5} is found by using Campbell's theorem \cite{Stochastic_Baccelli10_2} and the definition of $\gamma$. Finally, since the final result is independent of $\lambda_0$, then we can conclude that $ \lim\limits_{\lambda \rightarrow \infty} \frac{1}{\lambda} \sum\limits_{r_i \in \Phi \setminus B(0,r_o)}L(r_i) g_i=\lim\limits_{k \rightarrow \infty} \frac{1}{k\lambda_0} \sum\limits_{r_i \in \tilde{\Phi} \setminus B(0,r_o)}L(r_i) g_i=2\pi \gamma$, which proves the lemma.
\end{proof}
\end{lemma}

Next, we will utilize the expressions and tools of this section to derive the scaling laws for the different MIMO configurations, starting with MISO networks.
\section{Scenario A: MISO Networks with No BS Cooperation}\label{Sec:MISO}
For this scenario, we focus on the case where we have multi-antenna BSs and single-antenna users. In this case, eigenbeamforming simplifies to  maximum ratio transmission (MRT), i.e., $\bp_{i}=\frac{\bh_{i,i}}{||\bh_{i,i}||}$ \cite{Foundations_Heath18}, where $\bh_{i,j}\in \mathbb{C}^{N_t \times 1}$ is the channel between the $i^{\rm th}$ BS and the $j^{\rm th}$ user. In this case, the conditional SINR in \eqref{Eq:SINR_gen} can be written as
\begin{align}\label{Eq:SIMODigSINR}
   {\rm SINR}(\lambda)&=\frac{L(r_0)||\bh_{0,0}||^2}{\sum\limits_{r_i\in \Phi \setminus B(0,r_0)} L(r_i)\frac{\bh^{*}_{0}\bh_{i,0}\bh^{*}_{i,0}\bh_{0}}{||\bh_{0,0}||^2}+\sigma^2 }=\frac{L(r_0)\tilde{g}}{\sum\limits_{r_i\in \Phi \setminus B(0,r_0)} L(r_i)g_i+\sigma^2 },
\end{align}
where $\tilde{g}$ is a Gamma distributed random variable with shape $N_t$ and unit rate, i.e., $\tilde{g}\sim \Gamma(N_t,1)$, and $g_i, \forall i \in \{1, 2, \cdots\},$ are i.i.d. unit mean exponential random variables independent of $\tilde{g}$ \cite{Downlink_Dhillon13}.

\subsection{Scaling Laws}
Before delving into the analysis, it is important to recall that in the single antenna case, the conditional and the mean SINR drop to zero and the conditional and the mean ASE approach a finite constant as $\lambda \rightarrow \infty$ \cite{A_AlAmmouri19}.  For the multi-antenna case, one can use the SINR expression in \eqref{Eq:SIMODigSINR} to find the asymptotic SINR scaling laws as summarized in the next theorem.
\begin{theorem}\label{Th:SINRDig}
For the MISO case with $N_t(\lambda)$ antennas and a physically feasible path loss model, the conditional SINR has the following scaling law: $\lim\limits_{\lambda \rightarrow \infty}\frac{\lambda}{N_t(\lambda)}{\rm SINR (\lambda)}=\frac{L_0}{2 \pi \gamma}$ a.s., which is a finite constant. Equivalently:
\begin{enumerate}[{1}.1:]
    \item If $\lim\limits_{\lambda \rightarrow \infty}\frac{\lambda}{N_t(\lambda)}=\infty$, then ${\rm SINR}(\lambda) \rightarrow 0$ a.s. at a scale $\frac{\lambda}{N_t(\lambda)}$.
     \item If $\lim\limits_{\lambda \rightarrow \infty}\frac{\lambda}{N_t(\lambda)}=c\in \mathbb{R}^{*}_{+}$, then ${\rm SINR}(\lambda) \rightarrow \frac{L_0}{2 \pi \gamma c}$ a.s.
      \item If $\lim\limits_{\lambda \rightarrow \infty}\frac{\lambda}{N_t(\lambda)}=0$, then ${\rm SINR}(\lambda) \rightarrow \infty$ a.s. at a scale $\frac{N_t(\lambda)}{\lambda}$.
\end{enumerate}

If the path loss function also satisfies the conditions in Assumption 1, then the mean SINR scales similar to the conditional SINR asymptotically, i.e., 
\begin{align}
\lim\limits_{\lambda \rightarrow \infty} \mathbb{E}\left[\frac{\lambda}{N_t(\lambda)}{\rm SINR(\lambda)}\right]= \frac{L_0}{2 \pi \gamma}.\end{align}
\begin{proof}
Refer to Appendix \ref{app:SINRDig}.
\end{proof}
\end{theorem}

The proof of this theorem follows from the next two lemmas which we prove in the Appendix.

Hence, based on Theorem \ref{Th:SINRDig}.1, scaling the number of antennas sub-linearly with the density does not prevent the SINR from dropping to zero for high BS densities. The turning point happens when the number of antennas is scaled linearly with the density. In this case, the SINR approaches a finite constant which is desirable since it guarantees a certain QoS or throughput for the users in the dense regime. This roughly means that we can restore the SINR-invariance  property \cite{A_Andrews11} in dense networks by this scaling, under the assumption of a rich scattering environment we discussed before. For the ASE, the results are given in Theorem \ref{Th:ASEDig}.
\begin{theorem} \label{Th:ASEDig}
For the MISO case with $N_t(\lambda)$ antennas and a physically feasible path loss model, the conditional ASE scales as $\lambda \log\left(1+\frac{N_t(\lambda)}{\lambda}\right)$. Specifically:
\begin{enumerate}[{2}.1:]
    \item If $\lim\limits_{\lambda \rightarrow \infty}\frac{\lambda}{N_t(\lambda)}=\infty$, then $\mathcal{E}(\lambda) \rightarrow \infty$ at a scale of $N_t(\lambda)$.
     \item If $\lim\limits_{\lambda \rightarrow \infty}\frac{\lambda}{N_t(\lambda)}=c\in \mathbb{R}^{*}_{+}$, then $\mathcal{E}(\lambda) \rightarrow \infty$ at a scale of $\lambda$.
     \item If $\lim\limits_{\lambda \rightarrow \infty}\frac{\lambda}{N_t(\lambda)}=0$, then $\mathcal{E}(\lambda) \rightarrow \infty$ at a scale  $\lambda\log\left(1+\frac{N_t(\lambda)}{\lambda}\right)$.
\end{enumerate}

If the path loss function also satisfies the conditions in Assumption 1, then the mean ASE, i.e., $\mathbb{E}\left[\mathcal{E}(\lambda)\right]$, has the same scaling laws as above.
\begin{proof}
Refer to Appendix \ref{app:ASEDig}.
\end{proof}
\end{theorem}

Theorem \ref{Th:ASEDig}.1 shows that although the SINR drops to zero if the number of antennas scales sub-linearly with the BS density, we still observe benefits from densifying the network in terms of the sum spatial throughput. This is because the density of the links (users) grows at a rate  faster than the decay of the SINR. Hence, although the throughput of each user tends to zero asymptotically, the sum throughput of all users still grows with densification. Moreover, Theorem \ref{Th:ASEDig}.2 shows that a linear scaling, which is required to maintain a non-zero SINR, leads to a linear growth of the ASE in dense networks. Overall, the last theorem shows that as long as the number of antennas scales positively with the BS density, the densification plateau can be avoided. 

\subsection{Numerical Example}\label{Sec:DigSimu}

\begin{figure}[t]
        \centerline{\includegraphics[width= 4in]{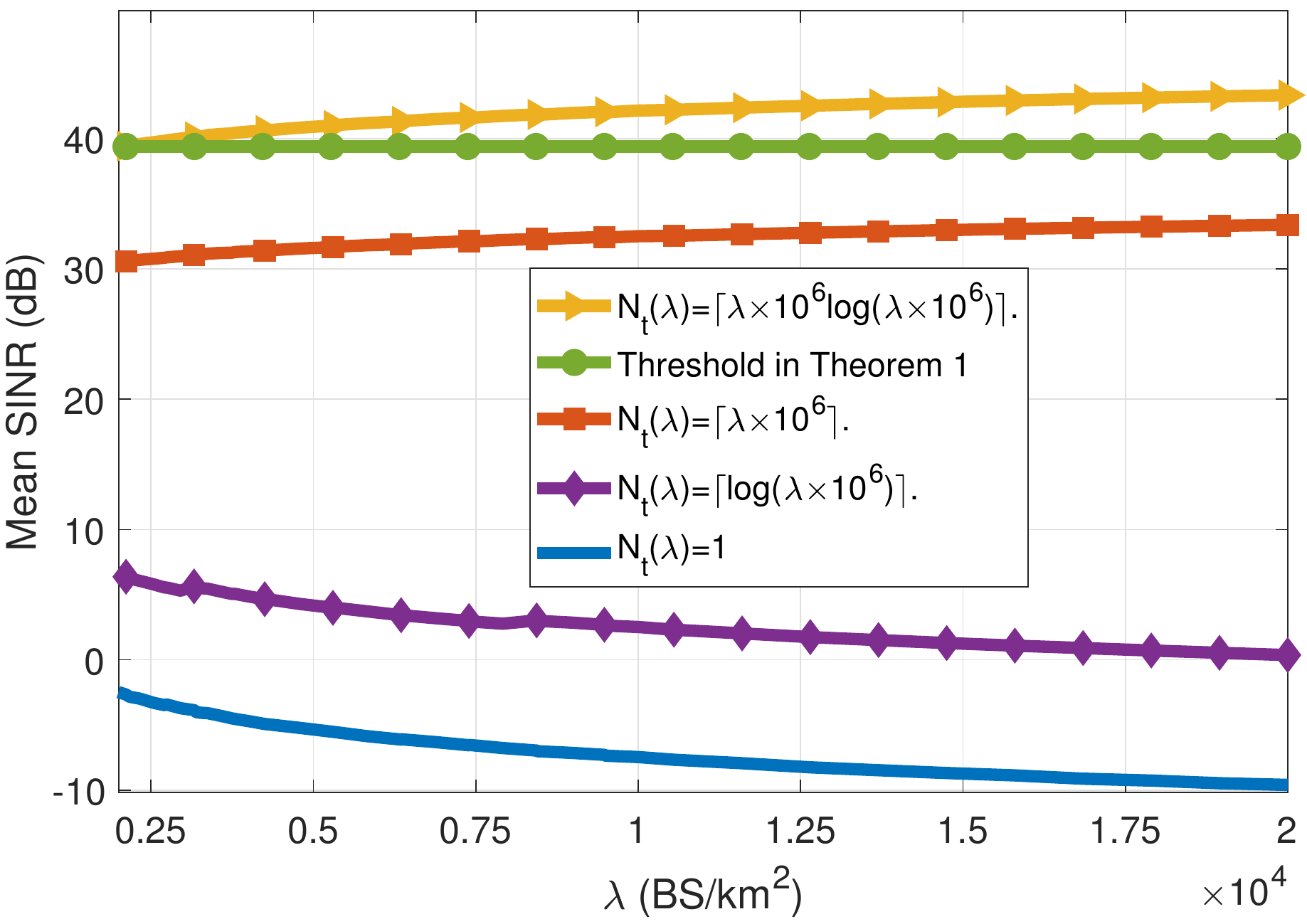}}
		\caption{\, Average SINR vs the BS density $\lambda$ for different scaling of the number of antennas for the MISO scenario.}
		\label{fig:SINR_MISO}
\end{figure}

\begin{figure}[t]
\centering
		\begin{subfigure}{.5\textwidth}				\centerline{\includegraphics[width= 3.15in]{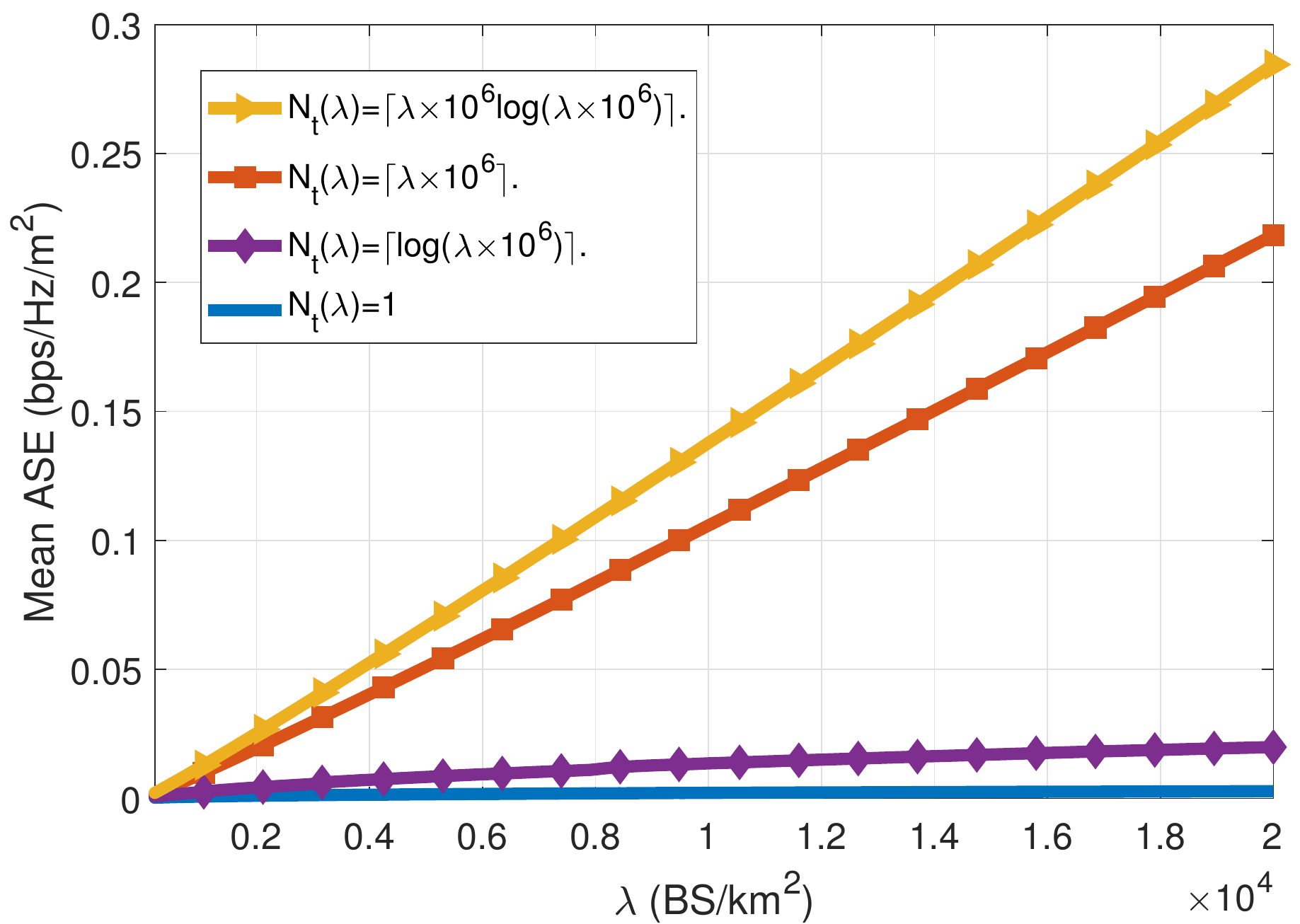}}
		\caption{\, Average ASE.}
		\label{fig:ASE}
		\end{subfigure}%
		\begin{subfigure}{.5\textwidth}
        \centerline{\includegraphics[width= 3.4in]{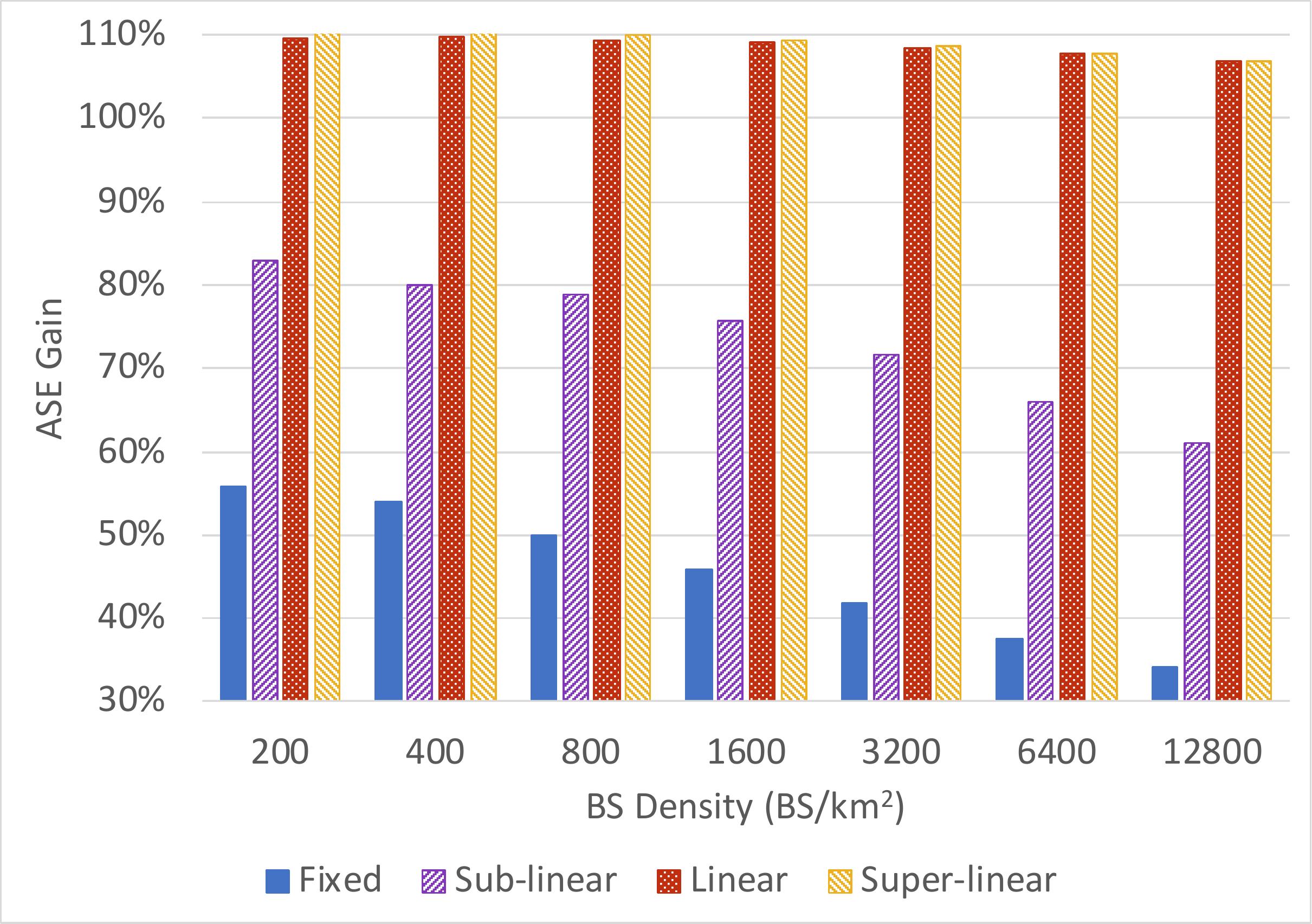}}
		\caption{\, Average ASE relative gain.}
		\label{fig:ASEGain_MISO}
		\end{subfigure}
		\caption{Average ASE vs the BS density $\lambda$ for different scaling of the number of antennas for the MISO scenario.}
		\label{fig:ASE_MISO}
\end{figure}

We start this section by verifying our derived scaling laws using independent and realistic system level simulations. The simulation uniformly drops BSs in a $20\times20$ km$^2$ region according to the desired density. Then the SINR is evaluated for a user located at the origin. The results were averaged over $10^{4}$ runs. Unless otherwise stated, the noise power is set to $\sigma^2=-70$dBm and the path loss is given by $L(r)=\exp(-\eta r^{-\kappa})$, with $\eta=0.9$ and $\kappa=0.52$. These values were picked since it was shown in \cite{SINR_AlAmmouri17} that using these parameters, the stretched exponential function accurately captures the path loss in dense urban networks based on the measurements provided in \cite{A_Franceschetti04}. The simulation results are shown in Fig. \ref{fig:SINR_MISO} and Fig. \ref{fig:ASE_MISO}. Fig. \ref{fig:SINR_MISO} shows the scaling of the mean SINR with the BS density for different scaling rates of the number of antennas; super-linear, linear, sub-linear, and constant (single antenna). We also include the asymptotic value for the linear scaling case given in Theorem \ref{Th:SINRDig}.2. The curves agree with and verify the derived scaling laws. Precisely, the figure shows that the SINR decreases with the density for the single antenna case, which was proven in \cite{A_AlAmmouri19}, and also when the number of antennas is scaled sub-linearly with the density, which we proved in Theorem \ref{Th:SINRDig}. 

The figure also shows that a linear scaling of the number of antennas with the BS density is required to prevent the SINR from dropping to zero. 
Fig. \ref{fig:SINR_MISO} also highlights the diminishing SINR gains we get by densification for the linear scaling case, although the limiting value is approached at very high densities, i.e., at densities much higher than $2\times10^4$ BSs/km$^2$. For example, to get 1dB SINR gain at a BS density of $3000$ BS/km$^2$, we need to increase the density to around $7000$ BS/km$^2$, and to get another 1dB gain, we need to densify the network to $15000$ BS/km$^2$. Practically, densities of $3000$, $7000$, and $15000$ BSs/km$^2$ correspond to a BS every $18$m, $12$m, and $8$m, respectively, assuming a uniform square grid model. Hence, improving the SINR through densification while scaling the number of antennas is not very attractive when the network is already dense. However, it still ensures a non-zero SINR which is our main objective, unlike the single-antenna case, where the SINR decreases to zero.

Moving to the mean ASE, Fig. \ref{fig:ASE}  illustrates the scaling laws derived in Theorem \ref{Th:ASEDig} and shows the high gains of densification with antenna scaling in terms of the network throughput. More specifically, Fig. \ref{fig:ASEGain_MISO} shows the ASE gain we get by doubling the BS density, where the gain at 200 BS/km$^2$ is relative to 100 BS/km$^2$. Note that although it was proven in \cite{A_AlAmmouri19} that the mean ASE saturates to constant in the limit $\lambda \rightarrow \infty$ for the single-antenna case, the figure shows that this limit is approached for very high BS densities that are not practical. Nevertheless, the figure highlights the diminishing gains we get by densifying the network. These diminishing gains are a result of using a physically feasible path loss model, since for the unbounded power-law model, it was proven that the ASE scales linearly with the BS density \cite{A_Andrews11}. The figure also highlights the linear scaling of the ASE when the number of antennas is scaled linearly with the BS density. 

\section{Scenario B: SIMO Networks with No Interference Cancellation}\label{Sec:SIMO}
In this scenario, we focus on SIMO networks, where we have single-antenna BSs and multi-antenna users. Hence, eigenbeamforming simplifies to maximum ratio combining (MRC), i.e., $\bu_i=\bh_{i,i}$ \cite{Foundations_Heath18}. Based on this, the SINR in \eqref{Eq:SINR_gen} can be written as
\begin{align}
   {\rm SINR}(\lambda)&=\frac{L(r_0)||\bh_{0,0}||^4}{\sum\limits_{r_i\in \Phi \setminus B(0,r_0)} L(r_i)\bh^{*}_{0}\bh_{i,0}\bh^{*}_{i,0}\bh_{0}+{||\bh_{0,0}||^2}\sigma^2 }=\frac{L(r_0)\tilde{g}}{\sum\limits_{r_i\in \Phi \setminus B(0,r_0)} L(r_i)g_i+\sigma^2 },\label{Eq:SINR_SIMO_Sc1}
\end{align}
where $\tilde{g}$ is Gamma distributed with shape $N_r(\lambda)$ and unit rate, i.e., $\tilde{g}\sim \Gamma(N_r(\lambda),1)$, and $g_i, \forall i \in \{1, 2, \cdots\},$ are i.i.d. unit mean exponential random variables independent of $g_i$  \cite{A_Afify16}. This expression exactly matches the one for the MISO case given in \eqref{Eq:SIMODigSINR}, with the exception that the mean $\tilde{g}$ is $N_r(\lambda)$ in this case instead of $N_t(\lambda)$ in \eqref{Eq:SIMODigSINR}. Hence, all the scaling laws for the MISO case extend exactly to the SIMO case. 

\begin{corollary}
For the SIMO case with $N_r(\lambda)$ antennas, the conditional and the mean SINR have the same scaling laws as in Theorem \ref{Th:SINRDig} and the conditional and the mean ASE have the same scaling laws as in Theorem \ref{Th:ASEDig}.
\end{corollary}

\section{Scenario C: MIMO Networks with No BS Cooperation}\label{Sec:MIMO}
Now we go back to the original settings, with multi-antenna BSs and multi-antenna users. More specifically, we assume that the BSs and the users are equipped with $N_t(\lambda)$ and $N_r(\lambda)$ antennas, respectively. We further focus on the practical case, where $N_r(\lambda)\leq N_t(\lambda)$, i.e., $\lim\limits_{\lambda \rightarrow \infty}\frac{N_r(\lambda)}{N_t(\lambda)}=y\in [0,1]$, where $y=0$ includes the MISO case we discussed. The SINR in this case is given in \eqref{Eq:SINR_MIMO2}. Note that the dependency on $N_t(\lambda)$ and $N_r(\lambda)$ is captured through the distribution of $\phi_0^{2}$, which has been well-studied in \cite{Largest_Kang03}, but it does not have a simple form. Nevertheless, we can still derive the exact scaling laws of the conditional SINR and the ASE.

\begin{theorem}\label{th:SINR_MIMO}
For the MIMO case with $N_t(\lambda)$ transmit antennas, eigenbeamforming, a single data stream, a physically feasible path loss model, and $N_r(\lambda)$ receive antennas, $\lim\limits_{\lambda \rightarrow \infty}\frac{N_r(\lambda)}{N_t(\lambda)}=y\in [0,1]$, the conditional SINR has the following scaling law: $\lim\limits_{\lambda \rightarrow \infty}\frac{\lambda}{N_t(\lambda)}{\rm SINR (\lambda)}=\frac{L_0(1+\sqrt{y})^2}{2 \pi \gamma }$ a.s. and the conditional ASE has the same scaling laws as in Theorem \ref{Th:ASEDig} with $N_t(\lambda)$ antennas.
\begin{proof}
   Refer to Appendix \ref{app:SINR_MIMO}.
\end{proof}
\end{theorem}

Hence, interestingly, the scaling laws are agnostic to the number of receive antennas and it matches the scaling laws we derived for the MISO case. More specifically, increasing the number of receive antennas just changes the constant to which $\frac{\lambda {\rm SINR}(\lambda)}{N_t(\lambda)}$ saturates to, but does not change the scaling law. Different from the previous cases, we are unable to derive the exact scaling laws for the average SINR and the average ASE. This is because, to the best of our knowledge, the exact scaling of $\lim\limits_{N_t,N_r\rightarrow \infty}\mathbb{E}[\phi_{0}^2]$ is not known.  Nevertheless, we can derive bounds on the scaling laws as in the following corollary.
\begin{corollary}
For the MIMO case with $N_t(\lambda)$ transmit antennas, $N_r(\lambda)$ receive antennas, eigenbeamforming, a single data stream, a physically feasible path loss model that satisfies the requirements in Assumption 1,  and  $\lim\limits_{\lambda \rightarrow \infty}\frac{N_r(\lambda)}{N_t(\lambda)}=y\in [0,1]$, the average SINR scales faster than $\frac{N_t(\lambda)}{\lambda}$ and slower than $\frac{N_t(\lambda)N_r(\lambda)}{\lambda}$. The average ASE scales at least as the MISO case with $N_t(\lambda)$ antennas and at most as the MISO case with $N_t(\lambda)N_r(\lambda)$ transmit antennas.
\begin{proof}
    Refer to Appendix \ref{app:SINR_MIMO}.
\end{proof}
\end{corollary}

\begin{figure}[t]
\centering
		\begin{subfigure}{.5\textwidth}				\centerline{\includegraphics[width= 3.2in]{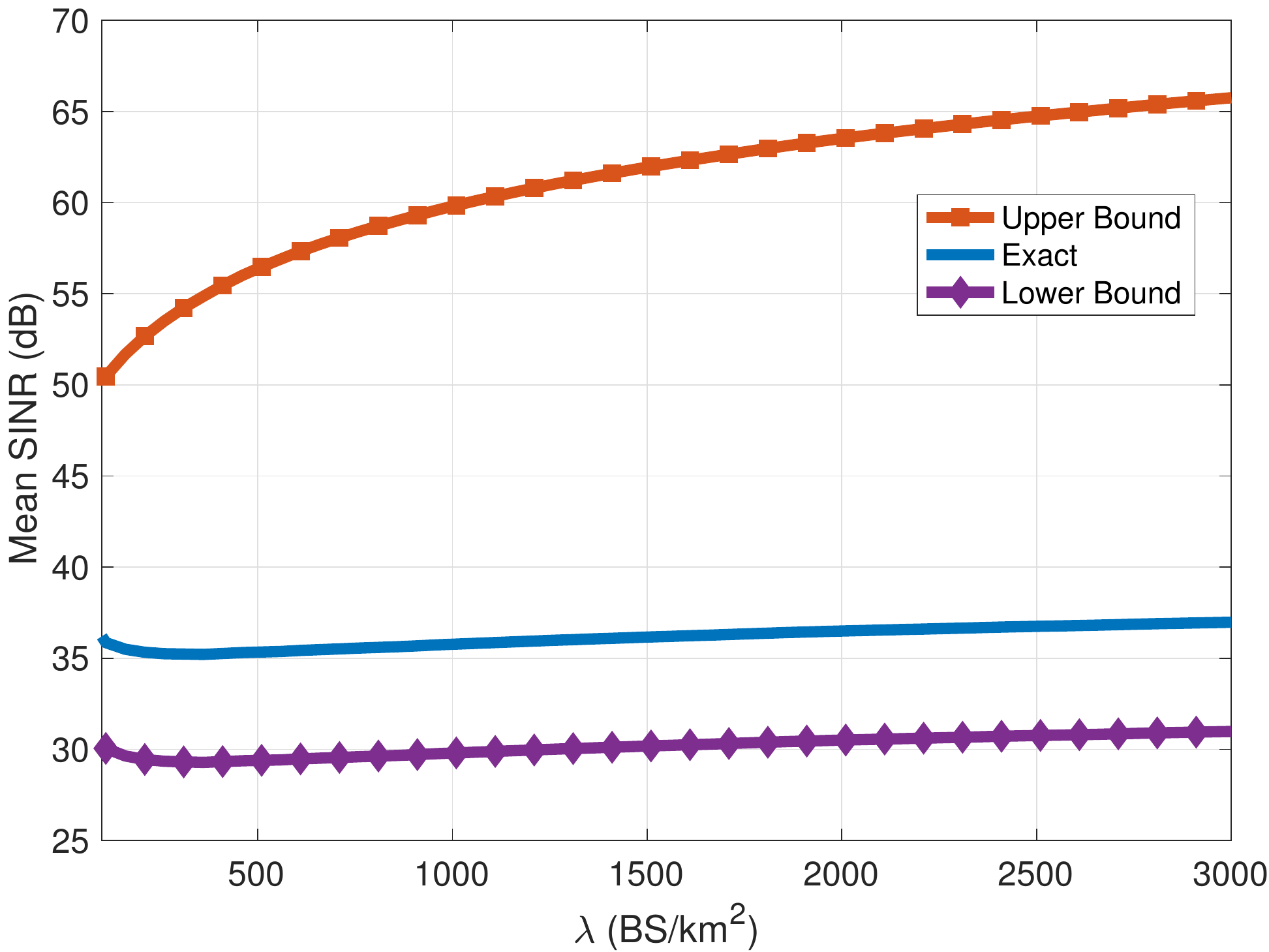}}
		\caption{\, Average SINR.}
		\label{fig:SINR_MIMO}
		\end{subfigure}%
		\begin{subfigure}{.5\textwidth}
        \centerline{\includegraphics[width= 3.2in]{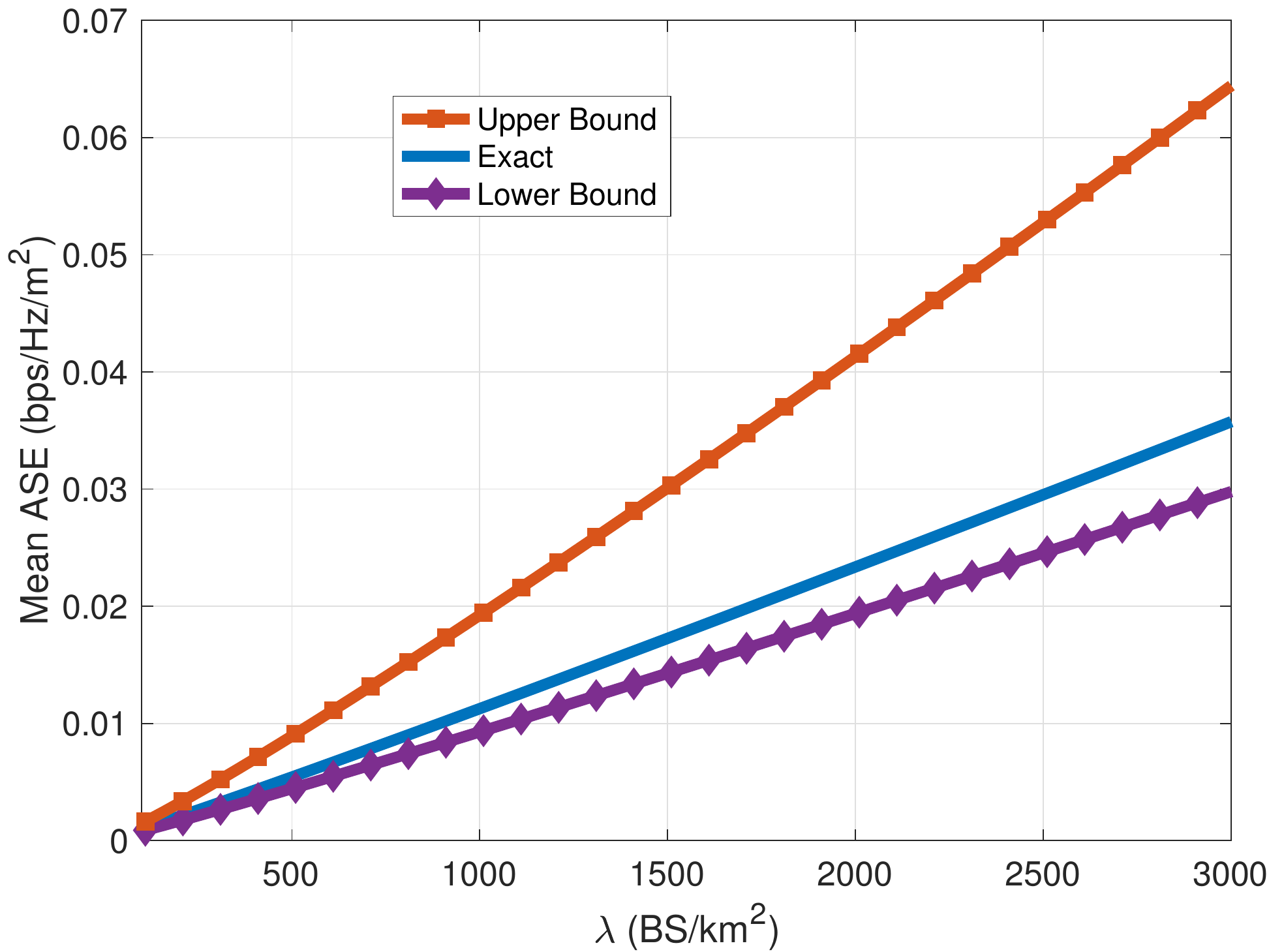}}
		\caption{\, Average ASE.}
		\label{fig:ASE_MIMO}
		\end{subfigure}
		\caption{Average ASE and SINR vs the BS density $\lambda$ assuming eigenbeamforming with $N_t(\lambda)=N_r(\lambda) =\lceil \lambda \times10^{6} \rceil$.}
		\label{fig:MIMO}
\end{figure} 

Hence, for the average SINR, the scaling law is at least similar to the MISO case with $N_t(\lambda)$ antennas and at most similar to the MISO case with $N_t(\lambda)N_r(\lambda)$ antennas. To observe the exact scaling, we use simulations and the results are shown in Fig. \ref{fig:MIMO}, assuming $N_t(\lambda)=N_r(\lambda)=N(\lambda) =\lceil \lambda \times10^{6} \rceil$. Starting with the mean SINR, Fig. \ref{fig:SINR_MIMO} shows that the mean SINR follows the same trend as the lower bound and not the upper bound. In other words, the mean SINR seems to scale as $\frac{N_t}{\lambda}$, which is a constant in this case. The results also show that the average ASE scales linearly with the BS density. More specifically, the ASE scaled as $\lambda$, as predicted by the lower bound, and not $\lambda \log_2(\lambda)$, predicted by the upper bound. It also matches the scaling law we derived for the conditional ASE.

\begin{conjecture}
For the MIMO case with $N_t(\lambda)$ transmit antennas, $N_r(\lambda)$ receive antennas, $\lim\limits_{\lambda \rightarrow \infty}\frac{N_r(\lambda)}{N_t(\lambda)}=y\in [0,1]$, eigenbeamforming, a single data stream, and a physically feasible path loss model that satisfies the requirements in Assumption 1, the average SINR scales as $\frac{N_t(\lambda)}{\lambda}$ and the average ASE scales as $\lambda\log\left(1+\frac{N_t(\lambda)}{\lambda}\right)$.
\end{conjecture}

\section{Scenario D: MISO Networks with Coordinated Beamforming}\label{Sec:MISOCoor}
 In the previous sections, we focused on the case where the BS only learns the channel to its user and designs its precoding vector according to that. In this section, we consider the case where the BSs jointly design their precoding vectors to mitigate the interference at their users, a scheme typically referred to as coordinated beamforming \cite{Spectral_Lee15}. More precisely, the cooperating BSs design their precoding vectors $\bv_i ,\forall i \in \{0,\cdots,K-1\}$, where $K$ is the number of cooperating BSs, such that the interference from this cluster of BSs is nullified at each user that is served by this cluster, while maximizing the desired signal power for each user. Such a construction of precoding vectors is always feasible as long as the cluster size is less than the number of antennas, i.e., $K\leq N_t(\lambda)$ \cite{Spectral_Lee15}.

Based on Lemma 1, canceling the interference from a finite number of BSs independently from the BS density does not help, since the interference term maintains the same scaling law. However, assuming that each BS is equipped with $N_t(\lambda)$ antennas, the precoding vectors can ideally be designed such that the interference from the $N_t(\lambda)-1$ closest interferers  is canceled at each user as we discussed.  Based on this, the SINR at the tagged user is \cite{Spectral_Lee15}
\begin{align}\label{Eq:SIMODigSINRCan}
   {\rm SINR}(\lambda)&=\frac{L(r_0)\tilde{g}}{\sum\limits_{r_i\in \Phi \setminus B(0,r_{N_t(\lambda)})} L(r_i)g_i+\sigma^2 },
\end{align}
where $r_{N_t(\lambda)}$ is the distance to the $N_t(\lambda)$ closest BS while $\tilde{g}$ and $g_i, \ \forall i \in \{1,2, \cdots \},$ have the same distributions as in \eqref{Eq:SIMODigSINR}. Note that we have neglected an important issue that comes with coordinated beamforming, which is how to form these cooperating clusters. In other words, we assumed that the closest BSs around the user at the origin are cooperating, but some of these might be shared with other clusters. Hence, the performance of the model as presented above, can be considered as an upper-bound on the actual model that accounts for the issue of BSs clustering. For more information on this issue, refer to \cite{Spectral_Lee15,Cooperative_Park16,A_Baccelli15}. Based on this, the SINR scaling laws are given in the following.

\begin{corollary}\label{Th:SINRDigCan}
For the MISO case with $N_t(\lambda)$ antennas, a physically feasible path loss model, and with coordinated beamforming such that the interference from the closest $N_t(\lambda)-1$ interfering BSs is perfectly canceled, the conditional SINR has the following scaling laws.
\begin{enumerate}[{4}.1:]
    \item If $\lim\limits_{\lambda \rightarrow \infty}\frac{\lambda}{N_t(\lambda)}=\infty$, then $\lim\limits_{\lambda \rightarrow \infty}\frac{\lambda}{N_t(\lambda)}{\rm SINR (\lambda)}=\frac{L_0}{2 \pi \gamma}$ a.s.
     \item If $\lim\limits_{\lambda \rightarrow \infty}\frac{\lambda}{N_t(\lambda)}=c>\frac{L_0}{2 \pi \gamma}$, then
\end{enumerate}
\begin{align}
   \frac{L_0}{2 \pi c \gamma}\leq \lim\limits_{\lambda \rightarrow \infty}{\rm SINR (\lambda)} \leq \frac{L_0}{2 \pi c \gamma-L_0}.
\end{align}

Accordingly, the conditional ASE has the same scaling laws as in Theorem~\ref{Th:ASEDig} for these two cases.
\begin{proof}
Refer to Appendix \ref{app:Can_1}.
\end{proof}
\end{corollary}

Hence, based on Corollary \ref{Th:SINRDigCan}.1, canceling the interference from the $N_t(\lambda)-1$ interferers does not improve the scaling law when the number of antennas is scaled sub-linearly with the BS density. In fact, even the slope does not change. The intuition behind this result is that the out-of-cluster interference grows faster than the in-cluster interference, which means that, asymptotically, the out-of-cluster interference dominates and converges to the total interference from all BSs. Hence, we can conclude that in this case, coordinated beamforming does not yield any benefits in terms of the asymptotic conditional SINR. 

\begin{figure}[t]
\centering
		\begin{subfigure}{.5\textwidth}
			\centerline{\includegraphics[width= 3.2in]{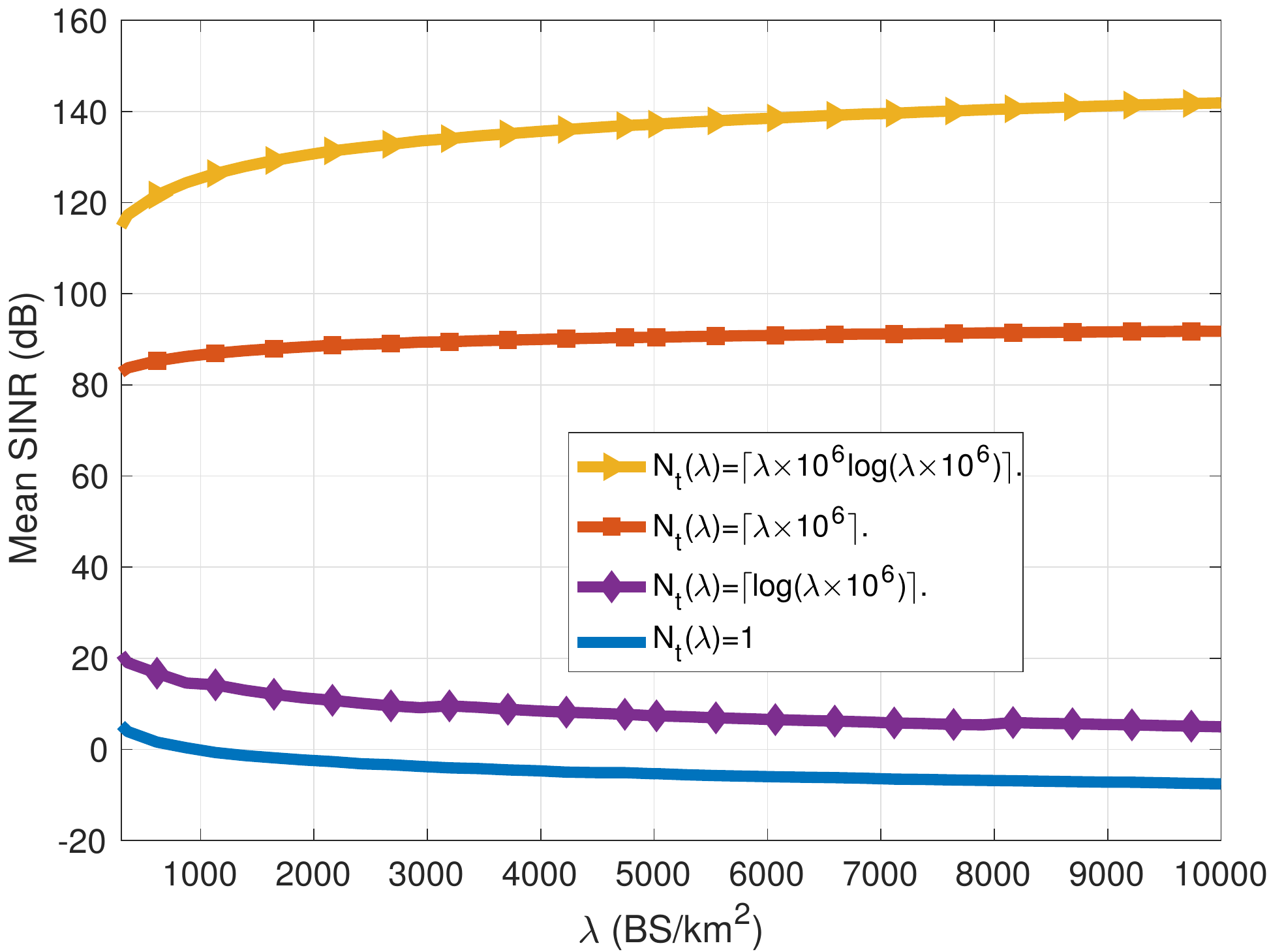}}
		\caption{\, Average SINR.}
		\label{fig:SINR_MISOCoor}
		\end{subfigure}%
		\begin{subfigure}{.5\textwidth}
        \centerline{\includegraphics[width= 3.2in]{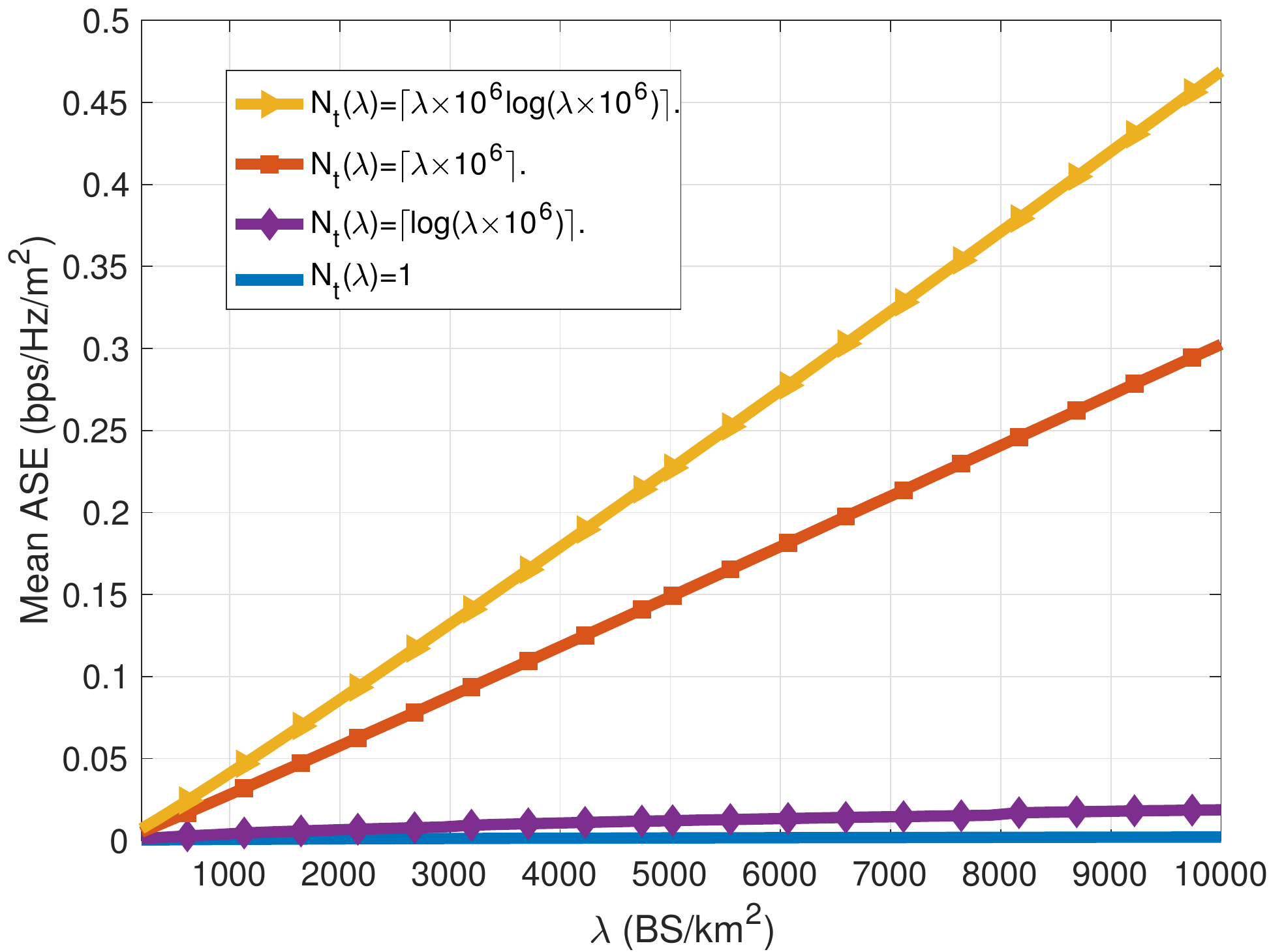}}
		\caption{\, Average ASE.}
		\label{fig:ASE_MISOCoor}
		\end{subfigure}
		\caption{Average ASE and SINR vs the BS density $\lambda$ for different scaling of the number of antennas for MISO scenario with coordinated beamforming.}
		\label{fig:MISOCoor}
\end{figure}

For the linear scaling case, the analysis becomes tricky, and hence, we focus on the special case where $c>\frac{L_0}{2 \pi \gamma}$ for simplicity, but we will verify by simulations that the same observation holds even if this condition is violated. More precisely, the previous corollary shows that even in the linear scaling case, the scaling law also does not change. Hence, although BS cooperation yields benefits for small densities in cellular networks, it does not improve the scaling laws in the ultradense regime when $\lambda \rightarrow \infty$. An example is shown in Fig. \ref{fig:MISOCoor}, where the simulation setup is similar to the one in Section \ref{Sec:DigSimu}, with the exception that the interference is perfectly canceled from the $N_t(\lambda)-1$ closest interferers. The results verify the last corollary and show that the scaling laws are similar to the ones in Fig. \ref{fig:SINR_MISO} and Fig. \ref{fig:ASE} for the MISO case without BS cooperation. Hence, we have the following conjecture.
\begin{conjecture}
For the MISO case with $N_t(\lambda)$ antennas, a physically feasible path loss model, and with coordinated beamforming such that the interference from the closest $N_t(\lambda)-1$ interfering BSs is perfectly canceled, the mean SINR scales as $\frac{N_t(\lambda)}{\lambda}$ and the mean ASE scales $\lambda \log_2\left(1+\frac{N_t(\lambda)}{\lambda}\right)$.
\end{conjecture}

Before wrapping up this section, we want to point that the expression in \eqref{Eq:SIMODigSINRCan} is similar to the SINR in a SIMO network, where the interference is perfectly cancelled from the $N_t(\lambda)-1$ closest interferes which was considered in \cite{Spectral_Lee16} for ad hoc networks. Hence, the scaling laws in Corollary \ref{Th:SINRDigCan} extend to this case as well.

\section{Discussion}\label{Sec:Discussion}
In this section, we provide further insights, discussions, comparisons, and possible future extensions of this work. All the scaling laws we derived are summarized in Table \ref{tab:Summary}.
\subsection{Practical implications}
First, recall that for the single antenna scenario, the SINR approaches zero and the ASE saturates to a non-zero finite constant in the limit of $\lambda \rightarrow \infty$ as shown in \cite{A_AlAmmouri19}. Hence, the per-user throughput, i.e., $\mathbb{E}[\log_2(1+{\rm SINR})]$, also approaches zero in the dense regime. The intuition behind this result is that although the per-user throughput is tiny, the sum throughput is still not negligible due to the high density of links. In fact, it was proved in \cite{A_AlAmmouri19} that if one would assume a minimum operational SINR $\theta$, such that if the received SINR is below this threshold, the packets are declared undecodable and discarded, then even the sum ASE approaches zero in this case. Hence, neither the per-user throughput nor the ASE benefit from densification after certain finite BS density.

Our results in this work suggest that network densification, along with scaling the number of antennas, is a sustainable way for cellular operators to cope with the massive increase in data demands, which is one of the main motivations behind the small cells technology \cite{What_Andrews14}. More precisely, scaling the number of antennas linearly with the BS density ensures a non-zero SINR and per-user throughput, at the same time, the network can support more devices which linearly increase the sum throughput. Hence, in scenarios where massive connectivity is expected, like in urban cellular networks or massive internet-of-things devices, these results show a durable solution for the massive connectivity challenge. Note that even if a minimum SINR threshold $\theta>0$ is imposed for successful transmission, we still observe the linear scaling of the ASE if we design $\theta$ to be larger than the constant the SINR saturates to, unlike the single-antenna case, where the ASE was proven to drop to zero in this case regardless of $\theta$.

\begin{table}[t]
\centering
\begin{tabular}{|c|c|c|c|}
\hline	\rowcolor[HTML]{EFEFEE} 
Scenarios &
  Antenna Scaling &
  SINR Scaling &
  ASE Scaling \\ \hline
Scenario A &
  $N_t(\lambda)$ positively scales with $\lambda$ &
  $\Theta\left(\frac{N_t(\lambda)}{\lambda}\right)$ &
  $\Theta\left(\lambda\log\left(1+\frac{N_t(\lambda)}{\lambda}\right)\right)$ \\ \hline
Scenario B&
  $N_r(\lambda)$ positively scales with $\lambda$ &
  $\Theta\left(\frac{N_r(\lambda)}{\lambda}\right)$ &
  $\Theta\left(\lambda\log\left(1+\frac{N_r(\lambda)}{\lambda}\right)\right)$ \\ \hline
Scenario C&
  $N_t(\lambda)$ positively scales with $\lambda$ with $N_r(\lambda)\leq N_t(\lambda)$ &
  $\Theta\left(\frac{N_t(\lambda)}{\lambda}\right)$ &
  $\Theta\left(\lambda\log\left(1+\frac{N_t(\lambda)}{\lambda}\right)\right)$ \\ \hline
Scenario D &
  $\lim\limits_{\lambda\rightarrow \infty}\frac{N_t(\lambda)}{\lambda}\leq\frac{2 \pi \gamma}{L_0}$ &
  $\Theta\left(\frac{N_t(\lambda)}{\lambda}\right)$ &
  $\Theta\left(\lambda\log\left(1+\frac{N_t(\lambda)}{\lambda}\right)\right)$ \\ \hline
\end{tabular}%
\caption{Summary of the scaling laws in different scenarios. For scenarios C and D, the scaling laws were only proven for the conditional SINR and ASE and verified by simulations for the average SINR and ASE.}
\label{tab:Summary}
\end{table}

\subsection{Comparison with the ad hoc scenario}
Shifting our focus from cellular networks, we compare our results to the ones in \cite{Spectral_Lee16}, where the authors studied the scaling laws of the ASE in an ad hoc network with multi-antenna receivers and single antenna transmitters. To have a fair comparison, we need to change our model slightly by assuming that the transmitter is located at a distance $r_0$ that is independent of the set of interfering transmitters and their densities. In other words, the SINR in \eqref{Eq:SINR_SIMO_Sc1} becomes
\begin{align}
   {\rm SINR}(\lambda)&= \frac{L(r_0)\tilde{g}}{\sum\limits_{r_i\in \Phi } L(r_i)g_i+\sigma^2 },\label{Eq:SINR_SIMO_Ad}
\end{align}
where $r_0$ is independent of $\lambda$ and uniformly distributed between $1$ and $R\in\mathbb{R}_{+}$. This expression matches the one in \cite{Spectral_Lee16} with the exception that we assume a physically feasible path loss model while in \cite{Spectral_Lee16}, the standard unbounded path loss model is considered, i.e., $L(r)=r^{-\eta}$ with $\eta>2$. Hence, any difference in the scaling laws is only related to the path loss model.

It is straightforward to prove that the scaling laws we derived for the cellular case match exactly the scaling laws in the ad hoc case based on \eqref{Eq:SINR_SIMO_Ad}. Compared to \cite{Spectral_Lee16}, the scaling laws we derived are more optimistic. More precisely, assuming $N_t(\lambda)= \lambda^{\beta}$ and no interference cancellation, the ASE in \cite{Spectral_Lee16} scales as $\lambda \log(1+ \lambda^{\beta-\eta/2})$, which is a function of the path loss exponent $\eta>2$. Hence, the ASE drops to zero in the single-antenna case and  $N_t(\lambda)$ has to scale super-linearly with $\lambda$ to maintain a linear scaling of the ASE. In our case, the ASE scales as $\lambda \log(1+\lambda^{\beta-1})$, which is independent of the path loss function as long as it physically feasible. Hence, the ASE saturates to a non-zero constant in the single-antenna case and a linear scaling of the $N_t(\lambda)$ is sufficient to maintain the linear scaling of the ASE with densification. Moreover, assuming perfect interference cancellation from the nearest $N_t(\lambda)-1$ interferers, the ASE in \cite{Spectral_Lee16} was shown to scale as $\lambda \log(1+\lambda^{\frac{\alpha}{2}(\beta-1)})$, while we showed that interference cancellation does not change the scaling laws of the ASE.

The differences between these scaling laws enforce our argument in \cite{A_AlAmmouri19} that one should be careful regarding the choice of the path loss model while studying the scaling laws in wireless networks.

\subsection{Coordinated Beamforming Overheads}
Coordinated beamforming is known to improve the SINR, at least for small BS densities, but at the expense of cooperation overheads, since the channels to all the users have to be known by all the cooperating BSs to design their precoders. This imposes a trade-off between the SINR and the overheads, and hence, the number of cooperating BSs has to be optimized. A more general cooperation scheme  has been studied for cellular networks in \cite{Fundamental_Lozano13}, but under the assumption of single-antenna BSs. It was shown in \cite{Fundamental_Lozano13} that the number of cooperating BSs is always finite; a small number of cooperating BSs yields low SINRs, but large number of cooperating BSs yields low throughput due to the overheads. In our case, we showed that even when we neglected the overheads, coordinated beamforming does not yield any benefits in terms of the ASE and SINR scaling laws.  Overall, this means that by accounting for the cooperation overheads, coordinated beamforming leads to worse scaling laws asymptotically, and hence, we cannot rely on it to improve the scaling laws in dense cellular networks.
\subsection{Open Questions}
\begin{itemize}
    \item Spatially Correlated Channels: all the scaling laws in this work are derived under the i.i.d. assumption for the channels seen by each antenna, regardless of the number of antennas. As mentioned before, this means we are focusing on traditional cellular frequency bands, but not mmWave or THz bands, since the channel is known to be spatially sparse on these frequencies, and thus highly correlated. Analyzing these frequency bands is interesting, since a large antenna array is more feasible. Related to this, understanding how much diversity in the channels is needed to maintain these scaling laws is also interesting and was not discussed in this work.
    \item MIMO with Statistical Multiplexing: we focused in this work on the MIMO case with a single data stream, since our objective is to  maintain a non-zero SINR, at least, in dense network to harvest the linear gain in the ASE. However, if the number of receive antennas is sufficient to provide enough diversity in the received signal to maintain a non-zero power, then perhaps transmitting multiple data streams would be favorable in this case. To the best of our knowledge, the optimal number of data streams is not known in dense cellular networks and is of  great interest.
    \item Open-loop MIMO: we focused on eigenbeamforming, where the channel is perfectly known at the BS. This requires ever more uplink feedback and/or pilot symbol transmissions as the number of BS antennas grows. Hence, it would be interesting to analyze the scaling laws of open-loop MIMO, where the BS is unaware of the channel to its user \cite{Foundations_Heath18}. This case was considered in \cite{Scaling_Lee18} for ad hoc networks, but the scaling laws in the cellular network context are unknown.
\end{itemize}


\appendices
   
\section{Proof of Theorem \ref{Th:SINRDig}}\label{app:SINRDig}
The proof follows by the following two lemmas, which we prove in the sequel.
\begin{lemma}\label{Lem:Thm1_2}
For the MISO case with $N_t(\lambda)$ antennas and a physically feasible path loss model, the conditional SINR in \eqref{Eq:SIMODigSINR} has the following scaling law.
\begin{align}
    \lim\limits_{\lambda \rightarrow \infty} \frac{\lambda}{N_t(\lambda)} {\rm SINR}(\lambda)=\frac{L_0}{2 \pi \gamma} \ {\rm a.s.}
\end{align}
\end{lemma}
\begin{lemma}\label{Lem:Thm1_3}
    For the MISO case with $N_t(\lambda)$ antennas and a physically feasible path loss model that satisfies Assumption 1, the mean SINR has the following scaling law.
    \begin{align}\label{eq:Lemma2_1}
        \lim\limits_{\lambda \rightarrow \infty} \mathbb{E} \left[ \frac{\lambda}{N_t(\lambda)} {\rm SINR}(\lambda)\right]=\frac{L_0}{2 \pi \gamma}.
    \end{align}
\end{lemma}
\subsection{Proof of Lemma \ref{Lem:Thm1_2}}\label{app:Lem2}
First, note that the SINR in \eqref{Eq:SIMODigSINR} can be written as
\begin{align}\label{Eq:SIMODigSINR2}
   {\rm SINR}(\lambda)&=\frac{L(r_0)\sum\limits_{n=1}^{N_t(\lambda)}f_i}{\sum\limits_{r_i\in \Phi \setminus B(0,r_0)}g_i L(r_i)+\sigma^2 },
\end{align}
where $f_i, \ \forall i\in \{1,2,\cdots, N_t(\lambda)\}$, are i.i.d. exponentially distributed random variables with unit means, which follows from the decomposition of the gamma random variable into a sum of i.i.d. exponential random variables. Based on this, we are interested in studying $\lim\limits_{\lambda \rightarrow \infty} \frac{\lambda}{N_t(\lambda)} {\rm SINR}(\lambda)$. Given that $\frac{\sigma^2}{\lambda} \rightarrow 0$ and $L(r_0) \rightarrow L_0$ a.s. as $\lambda \rightarrow \infty$, we have
\begin{align}
\lim\limits_{\lambda \rightarrow \infty} \frac{\lambda}{N_t(\lambda)} {\rm SINR}(\lambda)
        &=\lim\limits_{\lambda \rightarrow \infty}\frac{ L(r_0) \frac{1}{N_t(\lambda)}\sum\limits_{i=1}^{N_t(\lambda)}f_i}{\frac{1}{\lambda}\sum\limits_{r_i\in \Phi \setminus B(0,r_0)}g_i L(r_i)+\frac{\sigma^2}{\lambda} }=\frac{L_{0}\mathbb{E}[f_0]}{2 \pi \gamma} =\frac{L_{0}}{2 \pi \gamma} \ {\rm a.s.}\notag,
\end{align}
where the result follows using the law of large numbers and Lemma \ref{Lem:Thm1_1}, which concludes the proof of the conditional SINR scaling laws. 

\subsection{Proof of Lemma \ref{Lem:Thm1_3}}\label{app:Lem3}
For this lemma, we are interested in the scaling laws $ \lim\limits_{\lambda \rightarrow \infty} \mathbb{E} \left[ \frac{\lambda}{N_t(\lambda)} {\rm SINR}(\lambda)\right]$, which can be found as follows:
\begin{align}
        \lim\limits_{\lambda \rightarrow \infty} \mathbb{E} \left[ \frac{\lambda}{N_t(\lambda)} {\rm SINR}(\lambda)\right]&= \lim\limits_{\lambda \rightarrow \infty}\mathbb{E}\left[\frac{\lambda}{N_t(\lambda)}\frac{L(r_0)\sum\limits_{n=1}^{N_t(\lambda)}f_i}{\sum\limits_{r_i\in \Phi \setminus B(0,r_0)}g_i L(r_i)+\sigma^2 }\right]\notag\\
        &=\lim\limits_{\lambda \rightarrow \infty}\mathbb{E}\left[\frac{\lambda L(r_0)}{\sum\limits_{r_i\in \Phi \setminus B(0,r_0)}g_i L(r_i)+\sigma^2 }\right]\label{Eq:Thm_4_0}\\
        &=\mathbb{E}\left[\lim\limits_{\lambda \rightarrow \infty}\frac{\lambda L(r_0)}{\sum\limits_{r_i\in \Phi \setminus B(0,r_0)}g_i L(r_i)+\sigma^2 }\right]=\frac{L_{0}}{2 \pi \gamma},\label{Eq:Thm_4_1}
    \end{align}
where \eqref{Eq:Thm_4_0} is found by averaging over $\sum\limits_{n=1}^{N_t(\lambda)}f_i$ which has a mean $N_t(\lambda)$ and \eqref{Eq:Thm_4_1} holds since the random variables $\frac{\lambda L(r_0)}{\sum\limits_{r_i\in \Phi \setminus B(0,r_0)}g_i L(r_i)+\sigma^2 }$ are uniformly integrable with-respect-to (w.r.t) $\lambda$  given that $L(\cdot)$ satisfies the conditions in Assumption 1 as shown in \cite{A_AlAmmouri19}. Consequently, we can swap the limit with the expectation \cite[Theorem 5.5.2]{Probability_Durrett10}. Finally, the last equality holds since $L(r_0) \rightarrow L_0$ a.s. and $\frac{\sigma^2}{\lambda}\rightarrow 0$ as $\lambda \rightarrow \infty$, and then by using Lemma \ref{Lem:Thm1_1}.

\section{Proof of Theorem \ref{Th:ASEDig}}\label{app:ASEDig}
The proof follows by the following two lemmas, which we prove in the sequel.
\begin{lemma}\label{Lem:Thm2_1}
    For the MISO case with $N_t(\lambda)$ antennas and a physically feasible path loss model, the conditional ASE has the following scaling laws. If $\lim\limits_{\lambda \rightarrow \infty}\frac{\lambda}{N_t(\lambda)}=\infty$, then
    $
        \lim\limits_{\lambda \rightarrow \infty} \frac{ \mathcal{E}(\lambda)}{N_t(\lambda)}=\frac{L_0}{2 \pi \gamma \ln(2)} \ {\rm a.s.},
    $
    and if $\lim\limits_{\lambda \rightarrow \infty}\frac{\lambda}{N_t(\lambda)}=c\in \mathbb{R}_{+}$, then 
      $ \lim\limits_{\lambda \rightarrow \infty} \frac{\mathcal{E}(\lambda)}{N_t(\lambda)}  =c \log_2\left(1+\frac{L_{0}}{2 \pi \gamma}\right)\  {\rm a.s.}, $
      and if $\lim\limits_{\lambda \rightarrow \infty}\frac{\lambda}{N_t(\lambda)}=0$, then 
      $
        \lim\limits_{\lambda \rightarrow \infty}   \frac{\mathcal{E}(\lambda)}{\lambda \log_2(1+N_t(\lambda))}=1 \ {\rm a.s.}
      $
\end{lemma}

\begin{lemma}\label{Lem:Thm2_2}
    For the MISO case with $N_t(\lambda)$ antennas and a physically feasible path loss model that satisfies the conditions in Assumption 1, the mean ASE has the same scaling laws as the conditional ASE. 
\end{lemma}

\subsection{Proof of Lemma \ref{Lem:Thm2_1}}\label{app:Lem4}
 For the sub-linear case, Theorem \ref{Th:SINRDig} shows that the SINR approaches $0$ a.s. Hence, 
\begin{align}\label{eq:Thm2_1}
  \lim\limits_{\lambda \rightarrow \infty} \frac{\lambda\log_2(1+{\rm SINR}(\lambda))}{N_t(\lambda)} &= \lim\limits_{\lambda \rightarrow \infty} \frac{\lambda{\rm SINR}(\lambda)}{N_t(\lambda)\ln(2)}=\frac{L_0}{2 \pi \gamma \ln(2)} \ {\rm a.s.},
\end{align}
where the last equality follows from Theorem \ref{Th:SINRDig}. For the case where $\lim\limits_{\lambda \rightarrow \infty}\frac{\lambda}{N_t(\lambda)}=c$, then also using Theorem \ref{Th:SINRDig},
\begin{align}\label{eq:Thm2_2}
    \lim\limits_{\lambda \rightarrow \infty} \frac{\lambda}{N_t(\lambda)} \log_2(1+{\rm SINR}(\lambda)) =c \log_2\left(1+\frac{L_{0}}{2 \pi \gamma}\right) \ {\rm a.s.}
\end{align}

 For the super-linear case, Theorem 1 shows that the SINR approaches infinity at a scale $\frac{N_t(\lambda)}{\lambda}$. Also, there exit a $\lambda_0>0$ such that $\frac{L_{0}}{2 \pi \gamma} \frac{N_t(\lambda)}{\lambda} - \epsilon \leq{\rm SINR (\lambda)}\leq \frac{L_{0}}{2 \pi \gamma} \frac{N_t(\lambda)}{\lambda} + \epsilon $ a.s. for all $\lambda \geq \lambda_0$  and for any $\epsilon>0$. Hence, pick $\epsilon<1$ and then
\begin{align}
  \lim\limits_{\lambda \rightarrow \infty}\frac{\log\left(1+ \frac{L_{0}}{2 \pi \gamma} \frac{N_t(\lambda)}{\lambda} - \epsilon\right)}{\log\left(\frac{N_t(\lambda)}{\lambda}\right)} \leq \lim\limits_{\lambda \rightarrow \infty}   \frac{\mathcal{E}(\lambda)}{\lambda \log_2(N_t(\lambda))}\leq \lim\limits_{\lambda \rightarrow \infty}\frac{\log\left(1+ \frac{L_{0}}{2 \pi \gamma} \frac{N_t(\lambda)}{\lambda} + \epsilon\right)}{\log\left(\frac{N_t(\lambda)}{\lambda}\right)},
\end{align}
and both of the LHS and RHS evaluates to $1$. This concludes the proof of the scaling laws of the conditional ASE.
\subsection{Proof of Lemma \ref{Lem:Thm2_2}}\label{app:Lem5}
We have the following bounds
\begin{align}
\mathbb{E} \left[ \lim\limits_{\lambda \rightarrow \infty} \frac{\lambda\log_2(1+ {\rm SINR}(\lambda))}{N_t(\lambda)}\right]  \leq  \lim\limits_{\lambda \rightarrow \infty} \mathbb{E} \left[ \frac{\lambda\log_2(1+ {\rm SINR}(\lambda))}{N_t(\lambda)}\right]\leq \lim\limits_{\lambda \rightarrow \infty}  \frac{\lambda\mathbb{E} \left[{\rm SINR}(\lambda)\right]}{N_t(\lambda)\ln(2)} ,\label{eq:BOUNDS}
\end{align}
where the lower bound follows from Fatou's lemma \cite{Real_Royden88} and the upper bound follows since $\log_2(1+ x)\leq \frac{x}{\ln(2)}, \ \forall x\geq0$. For the sub-linear case, the LHS in \eqref{eq:BOUNDS} is $\frac{L_0}{2 \pi \gamma \ln(2)}$ according to \eqref{eq:Thm2_1} and the RHS is also  $\frac{L_0}{2 \pi \gamma \ln(2)}$ according to \eqref{Eq:Thm_4_1}. For the linear case, the LHS in \eqref{eq:BOUNDS} is $c \log_2\left(1+\frac{L_{0}}{2 \pi \gamma}\right)$ according to \eqref{eq:Thm2_2} and an alternative bound to the RHS in \eqref{eq:BOUNDS} can be found using Jensen's inequality, i.e., $\mathbb{E} \left[ \log_2(1+{\rm SINR(\lambda)})\right]\leq\log_2(1+\mathbb{E} \left[ {\rm SINR(\lambda)}\right]) $ and then  according to \eqref{Eq:Thm_4_1}, this bound is also $c \log_2\left(1+\frac{L_{0}}{2 \pi \gamma}\right)$. Based on these arguments, the mean ASE scales as $N_t(\lambda)$ for the sub-linear and the linear case.

For the super-linear case, we need to show that $\lim\limits_{\lambda \rightarrow \infty} \mathbb{E} \left[ \frac{\lambda \log_2(1+{\rm SINR(\lambda)})}{\lambda \log_2\left(1+\frac{N_t(\lambda)}{\lambda}\right)}\right]$ saturates to $1$. Note that
\begin{align}
\mathbb{E}\left[\lim\limits_{\lambda \rightarrow \infty}\frac{\log(1+{\rm SINR(\lambda)})}{\log\left(\frac{N_t(\lambda)}{\lambda}\right)}\right]\leq\lim\limits_{\lambda \rightarrow \infty}\frac{\mathbb{E}\left[\log(1+{\rm SINR(\lambda)})\right]}{\log\left(\frac{N_t(\lambda)}{\lambda}\right)}\leq \lim\limits_{\lambda \rightarrow \infty}\frac{\log(1+\mathbb{E}\left[{\rm SINR(\lambda)}\right])}{\log\left(\frac{N_t(\lambda)}{\lambda}\right)}\notag,
\end{align}
where the first inequality follows from Fatou's lemma \cite{Real_Royden88} and second from Jensen's inequality. The LHS evaluates to 1 using Lemma \ref{Lem:Thm2_1} and the RHS also evaluates to 1 using the same steps used to prove Lemma \ref{Lem:Thm2_1}. This concludes the proof of the super-linear case and the proof of Theorem \ref{Th:ASEDig}.

\section{MIMO Networks}\label{app:SINR_MIMO}
  \subsection{Proof of Theorem \ref{th:SINR_MIMO}}
  
         We are interested in 
       \begin{align}
\lim\limits_{\lambda \rightarrow \infty} \frac{\lambda {\rm SINR}(\lambda)}{N_t(\lambda)}&=\lim\limits_{\lambda \rightarrow \infty}\frac{\frac{1}{N_t(\lambda)}L(r_0) \phi_{0}^2}{\frac{1}{\lambda}\sum\limits_{r_i\in \Phi \setminus B(0,r_0)}L(r_i)g_i+\frac{\sigma^2}{\lambda}}\\
&=\frac{L_0(1+\sqrt{y})^2}{2 \pi \gamma } \ {\rm a.s.},
\end{align}
where the last equality holds since as $\lambda \rightarrow \infty$, $L(r_0) \rightarrow L_0$ a.s., $\frac{\sigma^2}{\lambda}\rightarrow0$,  $\frac{1}{\lambda}\sum\limits_{r_i\in \Phi \setminus B(0,r_0)}L(r_i)g_i\rightarrow 2 \pi \gamma$ a.s. as shown in Lemma~\ref{Lem:Thm1_1}, and  $ \frac{\phi^{2}_0}{N_t(\lambda)}\rightarrow(1+\sqrt{y})^2$ a.s. according to \cite[Proposition 6.2]{Eigenvalues_Edelman89}, which states that the maximum eigenvalue of a $N_r \times N_t$ matrix with entries drawn from i.i.d. complex Gaussian distribution with zero mean and unit variance approaches $(1+\sqrt{y})^2 N_t$ a.s. as $N_t \rightarrow \infty$, where $y=\frac{N_r}{N_t} \in [0,1]$, which is sometimes referred to as the semicircle law of random matrices.

\subsection{Proof of Corollary 2}

 The proof relies on the following bounds, which are derived in \cite{Matrix_Horn12}.
     \begin{align}
\frac{||\bH_{0,0}||_{\rm  F}}{\min(N_t(\lambda),N_r(\lambda))}\leq\phi_{0}^2\leq ||\bH_{0,0}||_{\rm F},
\end{align}
where $||\cdot||_{\rm F}$ is the Frobenius norm \cite{Matrix_Horn12}, i.e., $||\bH_{0,0}||_{\rm F}=\sum\limits_{i=1}^{N_{r}(\lambda)}\sum\limits_{j=1}^{N_{t}(\lambda)}|h_{i,j}|^2$. Hence, the proof follows from similar steps as in the MISO case, since $|h_{i,j}|^2, \ \forall i,j \in \{ 1,2, \cdots\}$, are i.i.d. unit mean exponential random variables.

\section{Proof of Corollary 3}\label{app:Can_1}
The conditional SINR in \eqref{Eq:SIMODigSINRCan} can be written as
\begin{align}\label{Eq:SIMODigSINRCan3}
   {\rm SINR}(\lambda)&=\frac{L(r_0)\tilde{g}}{\sum\limits_{r_i\in \Phi} L(r_i)g_i-\sum\limits_{j=0}^{N_t(\lambda)-1}L(r_j)g_j+\sigma^2 }.
\end{align}

Based on Lemma \ref{Lem:Thm1_1}, $\lim\limits_{\lambda \rightarrow \infty}\frac{1}{\lambda}\sum\limits_{r_i\in \Phi}L(r_i)g_i=2 \pi \gamma$ a.s., moreover, if $N_t(\lambda)$ scales sub-linearly with $\lambda$, then $\lim\limits_{\lambda \rightarrow \infty}\frac{1}{\lambda}\sum\limits_{j=0}^{N_t(\lambda)-1}L(r_j)g_j$ is upper-bounded by

\begin{align}
\lim\limits_{\lambda \rightarrow \infty}\frac{1}{\lambda} L_0\sum\limits_{j=0}^{N_t(\lambda)-1}g_j=L_0\lim\limits_{\lambda \rightarrow \infty}\frac{N_t(\lambda)}{\lambda} \frac{1}{N_t(\lambda)}\sum\limits_{j=0}^{N_t(\lambda)-1}g_j=L_0\lim\limits_{\lambda \rightarrow \infty}\frac{N_t(\lambda)}{\lambda}=0,\end{align}
where the law of large numbers is used to get the final result. Hence, we can conclude that $\frac{\lambda}{N_t(\lambda)} {\rm SINR}(\lambda)=\frac{L_0}{2 \pi}$ a.s. in case $\frac{N_t(\lambda)}{\lambda}$ approaches zero following the same steps as in Appendix~\ref{app:Lem2}. For the linear case,  i.e., $N_t(\lambda)=\frac{\lambda}{c}$ with $c>\frac{L_0}{2 \pi \gamma}$, following the previous steps we have 
\begin{align}\lim\limits_{\lambda \rightarrow \infty}\frac{1}{\lambda}\sum\limits_{j=0}^{N_t(\lambda)-1}L(r_j)g_j\leq \frac{L_0}{c}. \end{align}

Since $\lim\limits_{\lambda \rightarrow \infty}\frac{1}{\lambda}\sum\limits_{r_i\in \Phi}L(r_i)g_i=2 \pi \gamma$ a.s. based on Lemma \ref{Lem:Thm1_1}, then with the extra condition of $c>\frac{L_0}{2 \pi \gamma}$ to
have a valid upper-bound, the result is proven. The lower bound follows from the case without interference cancellation given in Theorem \ref{Th:SINRDig}.

\bibliographystyle{IEEEtran}
\bibliography{AhmadRef}

\begin{thebibliography}{10}
\providecommand{\url}[1]{#1}
\csname url@samestyle\endcsname
\providecommand{\newblock}{\relax}
\providecommand{\bibinfo}[2]{#2}
\providecommand{\BIBentrySTDinterwordspacing}{\spaceskip=0pt\relax}
\providecommand{\BIBentryALTinterwordstretchfactor}{4}
\providecommand{\BIBentryALTinterwordspacing}{\spaceskip=\fontdimen2\font plus
\BIBentryALTinterwordstretchfactor\fontdimen3\font minus
  \fontdimen4\font\relax}
\providecommand{\BIBforeignlanguage}[2]{{%
\expandafter\ifx\csname l@#1\endcsname\relax
\typeout{** WARNING: IEEEtran.bst: No hyphenation pattern has been}%
\typeout{** loaded for the language `#1'. Using the pattern for}%
\typeout{** the default language instead.}%
\else
\language=\csname l@#1\endcsname
\fi
#2}}
\providecommand{\BIBdecl}{\relax}
\BIBdecl

\bibitem{Scaling_AlAmmouri20}
\BIBentryALTinterwordspacing
A.~AlAmmouri, J.~G. Andrews, and F.~Baccelli, ``Scaling laws of dense
  multi-antenna cellular networks,'' \emph{CoRR}, vol. abs/2001.05083, Jan.
  2020. [Online]. Available: \url{https://arxiv.org/abs/2001.05083}
\BIBentrySTDinterwordspacing

\bibitem{A_Andrews11}
J.~G. Andrews, F.~Baccelli, and R.~K. Ganti, ``A tractable approach to coverage
  and rate in cellular networks,'' \emph{IEEE Trans. on Communications},
  vol.~59, no.~11, pp. 3122--3134, Nov. 2011.

\bibitem{A_AlAmmouri19}
A.~{AlAmmouri}, J.~G. {Andrews}, and F.~{Baccelli}, ``A unified asymptotic
  analysis of area spectral efficiency in ultradense cellular networks,''
  \emph{IEEE Trans. on Info. Theory}, vol.~65, no.~2, pp. 1236--1248, Feb.
  2019.

\bibitem{Downlink_Zhang15}
X.~Zhang and J.~G. Andrews, ``Downlink cellular network analysis with
  multi-slope path loss models,'' \emph{IEEE Trans. on Communications},
  vol.~63, no.~5, pp. 1881--1894, May 2015.

\bibitem{Performance_Ding17}
M.~Ding and D.~L{\'o}pez-P{\'e}rez, ``Performance impact of base station
  antenna heights in dense cellular networks,'' \emph{IEEE Trans. on Wireless
  Communications}, vol.~16, no.~12, pp. 8147--8161, Dec. 2017.

\bibitem{Performance_Nguyen17}
V.~M. Nguyen and M.~Kountouris, ``Performance limits of network
  densification,'' \emph{IEEE Journal on Sel. Areas in Communications},
  vol.~35, no.~6, pp. 1294--1308, Jun. 2017.

\bibitem{SINR_AlAmmouri17}
A.~AlAmmouri, J.~G. Andrews, and F.~Baccelli, ``{SINR} and throughput of dense
  cellular networks with stretched exponential path loss,'' \emph{IEEE Trans.
  on Wireless Communications}, vol.~17, no.~2, pp. 1147--1160, Feb. 2018.

\bibitem{The_Gupta00}
P.~Gupta and P.~R. Kumar, ``The capacity of wireless networks,'' \emph{IEEE
  Trans. on Info. Theory}, vol.~46, no.~2, pp. 388--404, Mar. 2000.

\bibitem{Stochastic_Baccelli10_2}
F.~Baccelli and B.~B{\l}aszczyszyn, ``Stochastic geometry and wireless
  networks: Volume {II} applications,'' \emph{Foundations and Trends in
  Networking}, vol.~4, no. 1--2, pp. 1--312, 2010.

\bibitem{Stochastic_Haenggi12}
M.~Haenggi, \emph{Stochastic Geometry for Wireless Networks}.\hskip 1em plus
  0.5em minus 0.4em\relax Cambridge University Publishers, 2012.

\bibitem{Stochastic_ElSawy13}
H.~ElSawy, E.~Hossain, and M.~Haenggi, ``Stochastic geometry for modeling,
  analysis, and design of multi-tier and cognitive cellular wireless networks:
  A survey,'' \emph{IEEE Communications Surveys \& Tutorials}, vol.~15, no.~3,
  pp. 996--1019, Jul. 2013.

\bibitem{Transmission_Weber11}
S.~Weber and J.~G. Andrews, ``Transmission capacity of wireless networks,''
  \emph{Foundations and Trends in Networking}, vol.~5, no. 2-3, pp. 109--281,
  2011.

\bibitem{Transmission_Hunter08}
A.~M. {Hunter}, J.~G. {Andrews}, and S.~{Weber}, ``Transmission capacity of ad
  hoc networks with spatial diversity,'' \emph{IEEE Trans. on Wireless
  Communications}, vol.~7, no.~12, pp. 5058--5071, Dec. 2008.

\bibitem{Open_Louie11}
R.~{Louie}, M.~{McKay}, and I.~{Collings}, ``Open-loop spatial multiplexing and
  diversity communications in ad hoc networks,'' \emph{IEEE Trans. on Info.
  Theory}, vol.~57, no.~1, pp. 317--344, Jan. 2011.

\bibitem{Transmission_Vaze12}
R.~{Vaze} and R.~W. {Heath}, ``Transmission capacity of ad-hoc networks with
  multiple antennas using transmit stream adaptation and interference
  cancellation,'' \emph{IEEE Transactions on Information Theory}, vol.~58,
  no.~2, pp. 780--792, Feb. 2012.

\bibitem{Multi_Jindal10}
N.~{Jindal}, J.~G. {Andrews}, and S.~{Weber}, ``Multi-antenna communication in
  ad hoc networks: Achieving {MIMO} gains with {SIMO} transmission,''
  \emph{IEEE Trans. on Communications}, vol.~59, no.~2, pp. 529--540, Feb.
  2011.

\bibitem{Spatial_Huang12}
K.~{Huang}, J.~G. {Andrews}, D.~{Guo}, R.~W. {Heath}, and R.~A. {Berry},
  ``Spatial interference cancellation for multiantenna mobile ad hoc
  networks,'' \emph{IEEE Trans. on Info. Theory}, vol.~58, no.~3, pp.
  1660--1676, Mar. 2012.

\bibitem{Spectral_Lee16}
N.~Lee, F.~Baccelli, and R.~W. Heath, ``Spectral efficiency scaling laws in
  dense random wireless networks with multiple receive antennas,'' \emph{IEEE
  Trans. on Info. Theory}, vol.~62, no.~3, pp. 1344--1359, Mar. 2016.

\bibitem{Scaling_Lee18}
J.~{Lee}, N.~{Lee}, and F.~{Baccelli}, ``Scaling laws for ergodic spectral
  efficiency in {MIMO} poisson networks,'' \emph{IEEE Trans. on Info. Theory},
  vol.~64, no.~4, pp. 2791--2804, Apr. 2018.

\bibitem{Are_Lozano12}
A.~{Lozano} and N.~{Jindal}, ``Are yesterday-s information-theoretic fading
  models and performance metrics adequate for the analysis of today's wireless
  systems?'' \emph{IEEE Communications Magazine}, vol.~50, no.~11, pp.
  210--217, Nov. 2012.

\bibitem{Modeling_ElSawy16}
H.~ElSawy, A.~Sultan-Salem, M.~S. Alouini, and M.~Z. Win, ``Modeling and
  analysis of cellular networks using stochastic geometry: A tutorial,''
  \emph{IEEE Communications Surveys \& Tutorials}, vol.~19, no.~1, pp.
  167--203, Firstquarter 2017.

\bibitem{A_Andrews16}
J.~G. Andrews, A.~K. Gupta, and H.~S. Dhillon, ``A primer on cellular network
  analysis using stochastic geometry,'' \emph{arXiv preprint arXiv:1604.03183},
  2016.

\bibitem{Stochastic_Bartek18}
B.~B{\l}aszczyszyn, M.~Haenggi, P.~Keeler, and S.~Mukherjee, \emph{Stochastic
  geometry analysis of cellular networks}.\hskip 1em plus 0.5em minus
  0.4em\relax Cambridge University Press, 2018.

\bibitem{Downlink_Dhillon13}
H.~S. Dhillon, M.~Kountouris, and J.~G. Andrews, ``Downlink {MIMO} {HetNets}:
  Modeling, ordering results and performance analysis,'' \emph{IEEE Trans. on
  Wireless Communications}, vol.~12, no.~10, pp. 5208--5222, Oct. 2013.

\bibitem{A_Renzo14}
M.~{Di Renzo} and P.~{Guan}, ``A mathematical framework to the computation of
  the error probability of downlink {MIMO} cellular networks by using
  stochastic geometry,'' \emph{IEEE Trans. on Communications}, vol.~62, no.~8,
  pp. 2860--2879, Aug. 2014.

\bibitem{Stochastic_Renzo15}
M.~{Di Renzo} and W.~{Lu}, ``Stochastic geometry modeling and performance
  evaluation of {MIMO} cellular networks using the equivalent-in-distribution
  ({EiD})-based approach,'' \emph{IEEE Trans. on Communications}, vol.~63,
  no.~3, pp. 977--996, Mar. 2015.

\bibitem{A_Afify16}
L.~H. {Afify}, H.~{ElSawy}, T.~Y. {Al-Naffouri}, and M.~S. {Alouini}, ``A
  unified stochastic geometry model for {MIMO} cellular networks with
  retransmissions,'' \emph{IEEE Trans. on Wireless Communications}, vol.~15,
  no.~12, pp. 8595--8609, Dec. 2016.

\bibitem{Analysis_Tanbourgi15}
R.~{Tanbourgi}, H.~S. {Dhillon}, and F.~K. {Jondral}, ``Analysis of joint
  transmit–receive diversity in downlink {MIMO} heterogeneous cellular
  networks,'' \emph{IEEE Trans. on Communications}, vol.~14, no.~12, pp.
  6695--6709, Dec. 2015.

\bibitem{Empirical_Hata80}
M.~Hata, ``Empirical formula for propagation loss in land mobile radio
  services,'' \emph{IEEE Trans. on Veh. Technology}, vol.~29, no.~3, pp.
  317--325, Aug. 1980.

\bibitem{3GPP2017}
{3GPP TR 38.901}, in \emph{Study on channel model for frequencies from 0.5 to
  100 GHz ({R}elease 14)},
  \url{http://www.3gpp.org/ftp/Specs/archive/38series/38.901/38901-e20.zip},
  Sep. 2017.

\bibitem{Foundations_Heath18}
R.~W. Heath~Jr and A.~Lozano, \emph{Foundations of MIMO communication}.\hskip
  1em plus 0.5em minus 0.4em\relax Cambridge University Press, 2018.

\bibitem{Millimeter_Rappaport14}
T.~S. Rappaport, R.~W. Heath~Jr, R.~C. Daniels, and J.~N. Murdock,
  \emph{Millimeter wave wireless communications}.\hskip 1em plus 0.5em minus
  0.4em\relax Pearson Education, 2014.

\bibitem{The_Afify15}
L.~H. {Afify}, H.~{ElSawy}, T.~Y. {Al-Naffouri}, and M.~S. {Alouini}, ``The
  influence of {Gaussian} signaling approximation on error performance in
  cellular networks,'' \emph{IEEE Communications Letters}, vol.~19, no.~12, pp.
  2202--2205, Dec. 2015.

\bibitem{Influence_Giorgetti05}
A.~{Giorgetti} and M.~{Chiani}, ``Influence of fading on the {Gaussian}
  approximation for {BPSK} and {QPSK} with asynchronous cochannel
  interference,'' \emph{IEEE Trans. on Wireless Communications}, vol.~4, no.~2,
  pp. 384--389, Mar. 2005.

\bibitem{Area_Alouini99}
M.~S. Alouini and A.~J. Goldsmith, ``Area spectral efficiency of cellular
  mobile radio systems,'' \emph{IEEE Trans. on Veh. Technology}, vol.~48,
  no.~4, pp. 1047--1066, Jul. 1999.

\bibitem{Real_Royden88}
H.~Royden, \emph{Real Analysis}, ser. Mathematics and statistics.\hskip 1em
  plus 0.5em minus 0.4em\relax Macmillan, 1988.

\bibitem{A_Franceschetti04}
M.~Franceschetti, J.~Bruck, and L.~J. Schulman, ``A random walk model of wave
  propagation,'' \emph{IEEE Trans. on Antennas and Propagation}, vol.~52,
  no.~5, pp. 1304--1317, May 2004.

\bibitem{Largest_Kang03}
M.~Kang and M.-S. {Alouini}, ``Largest eigenvalue of complex {Wishart} matrices
  and performance analysis of {MIMO} {MRC} systems,'' \emph{IEEE Journal on
  Sel. Areas in Communications}, vol.~21, no.~3, pp. 418--426, Apr. 2003.

\bibitem{Spectral_Lee15}
N.~{Lee}, D.~{Morales-Jimenez}, A.~{Lozano}, and R.~W. {Heath}, ``Spectral
  efficiency of dynamic coordinated beamforming: A stochastic geometry
  approach,'' \emph{IEEE Trans. on Wireless Communications}, vol.~14, no.~1,
  pp. 230--241, Jan. 2015.

\bibitem{Cooperative_Park16}
J.~{Park}, N.~{Lee}, and R.~W. {Heath}, ``Cooperative base station coloring for
  pair-wise multi-cell coordination,'' \emph{IEEE Trans. on Communications},
  vol.~64, no.~1, pp. 402--415, Jan. 2016.

\bibitem{A_Baccelli15}
F.~Baccelli and A.~Giovanidis, ``A stochastic geometry framework for analyzing
  pairwise-cooperative cellular networks,'' \emph{IEEE Trans. on Wireless
  Communications}, vol.~14, no.~2, pp. 794--808, Feb. 2015.

\bibitem{What_Andrews14}
J.~G. Andrews \emph{et~al.}, ``What will {5G} be?'' \emph{IEEE Journal on Sel.
  Areas in Communications}, vol.~32, no.~6, pp. 1065--1082, Jun. 2014.

\bibitem{Fundamental_Lozano13}
A.~{Lozano}, R.~W. {Heath}, and J.~G. {Andrews}, ``Fundamental limits of
  cooperation,'' \emph{IEEE Trans. on Info. Theory}, vol.~59, no.~9, pp.
  5213--5226, Sep. 2013.

\bibitem{Probability_Durrett10}
R.~Durrett, \emph{Probability: theory and examples}.\hskip 1em plus 0.5em minus
  0.4em\relax Cambridge university press, 2010.

\bibitem{Eigenvalues_Edelman89}
A.~Edelman, ``Eigenvalues and condition numbers of random matrices,'' Ph.D.
  dissertation, Massachusetts Institute of Technology, 1989.

\bibitem{Matrix_Horn12}
R.~A. Horn and C.~R. Johnson, \emph{Matrix analysis}.\hskip 1em plus 0.5em
  minus 0.4em\relax Cambridge university press, 2012.

\end{thebibliography}
\vfill
\end{document}